%% file: 0_shocklet_PRF_arxive.tex
\documentclass[aps,PRE,groupedaddress,longbibliography]{revtex4-1}
\usepackage{graphicx,color}
\usepackage{epstopdf}
\usepackage{pgfplots}
\pgfplotsset{compat=newest}
\usetikzlibrary{plotmarks}
\usepackage{grffile}
\usepackage{amsmath}
\usepackage{natbib}

\usepackage{ulem}
\usepackage{color}

\usepackage{placeins}
\usepackage{textcomp}
\usepackage[euler]{textgreek}
\usepackage{xspace}
\usepackage{colortbl}
\usepackage[T1]{fontenc}
\usepackage{scalerel}
\usepackage{subcaption} 
\usepackage{footnote}
\usepackage{textcomp}
\usepackage{comment}
\usepackage{microtype} % text nicht über Rand
\usepackage{booktabs}

\newcommand{\drhodx}{$\frac{\partial \rho }{\partial x}$\xspace}
\newcommand{\drhody}{$\frac{\partial \rho }{\partial y}$ \xspace}
\newcommand{\NPG}{$- \nabla p$\ }
\newcommand{\myquote}[1]{``#1''}
\newcommand{\scales}{{\scaleto{s\mathstrut}{6pt}}}

\newcommand{\tdreiunddreissig}{$\tau = 5.81\ $}
\newcommand{\tdreiunddreissigX}{\tau = 5.81}
\newcommand{\tdreiunddreissigY}{$\tau = 5.81 $}

\newcommand{\tFuenfUndDreissig}{$\tau = 6.30\ $}

\newcommand{\tFuenfUndDreissigY}{$\tau = 6.30 $}

\newcommand{\tAchtUndZwanzig}{$\tau = 4.60\ $}
\newcommand{\tAchtUndZwanzigX}{\tau = 4.60 }
\newcommand{\tAchtUndZwanzigY}{$\tau = 4.60 $}

\newcommand{\tDreissig}{$\tau = 5.09\ $}
\newcommand{\tDreissigX}{\tau = 5.09}

\newcommand{\tSiebenUndZwanzigY}{$\tau = 4.36 $}

\newcommand{\tFuenfUndZwanzig}{$\tau = 3.88\ $}
\newcommand{\tFuenfUndZwanzigX}{\tau = 3.88 }
\newcommand{\tFuenfUndZwanzigY}{$\tau = 3.88 $}

\newcommand{\tZwanzig}{$\tau = 2.66\ $}

\newcommand{\tZwanzigY}{2.66}

\newcommand{\tEinundZwanzigY}{$\tau = 2.90 $}

\newcommand{\tZweiUndZwanzig}{$\tau = 3.15\ $}

\newcommand{\tZweiUndZwanzigY}{$\tau = 3.15 $}
\definecolor{dunkelgrau}{rgb}{0.87,0.87,0.87}
\definecolor{hellgrau}{rgb}{0.93,0.93,0.93}
\definecolor{sehrhellgrau}{rgb}{0.97,0.97,0.97}

\newcommand{\rk}[1]{\textcolor{black}{#1}}
\begin{document}
	\title{Dynamic Evolution of a Transient Supersonic Trailing Jet Induced by a Strong Incident Shock Wave}
	
	\author{Mohammad \surname{Rezay Haghdoost}}
	\affiliation{Laboratory for Flow Instabilities and Dynamics, Technische Universit\"at Berlin, Germany}
	\email{rezayhaghdoost@tu-berlin.de}
	
	\author{Daniel \surname{Edgington-Mitchell}}
	\affiliation{Department of Mechanical and Aerospace Engineering, Monash University, Melbourne, Australia}
	
	\author{Maikel \surname{Nadolski}}
	\affiliation{Freie Universit\"at Berlin, Berlin, Germany}
	
	\author{Rupert \surname{Klein}}
	\affiliation{Freie Universit\"at Berlin, Berlin, Germany}
	
	\author{Kilian\surname{ Oberleithner}}
	\affiliation{Laboratory for Flow Instabilities and Dynamics, Technische Universit\"at Berlin, Germany }
	
	\date{\today}
	
	\begin{abstract}
		
		The dynamic evolution of a highly underexpanded transient supersonic jet at the exit of a pulse detonation engine is investigated via high-resolution time-resolved schlieren and numerical simulations. Experimental evidence is provided for the presence of a second triple shock configuration along with a shocklet between the reflected shock and the slipstream, which has no analogue in a steady-state underexpanded jet. A pseudo-steady model is developed, which allows for the determination of the post-shock flow condition for a transient propagating oblique shock. This model is applied to the numerical simulations to reveal the mechanism leading to the formation of the second triple point. Accordingly, the formation of the triple point is initiated by the transient motion of the reflected shock, which is induced by the convection of the vortex ring. While the vortex ring embedded shock move essentially as a translating strong oblique shock, the reflected shock is rotating towards its steady state position. This results in a pressure discontinuity that must be resolved by the formation of a shocklet.
		
	\end{abstract}
	
	%\keywords{}
	
	\maketitle
	
	% References should be done using the \cite,~\ref, and \label commands

	\section{Introduction} \label{intro}
	
	Supersonic transient underexpanded compressible jets can be found in many applications such as rocket propulsion, shock tubes, pulse detonation engines, etc. The transient supersonic jet is also of interest in the field of safety and security management, e.g., in case of an accidental release of a gas from a high-pressure reservoir or volcanic blasts. The characterization of such a flow field has been the subject of research in some detail for many years. The first stage of the jet evolution is the well known shock-diffraction phenomenon, which has been investigated numerically, experimentally and analytically by many researchers \cite{skews1967shape,takayama1981formation,sun2003vorticity,takayama1981formation,bazhenova1984unsteady}. The next stage is the dynamic evolution of a highly transient supersonic trailing jet behind the leading shock, which has also received significant attention \cite{kleine2010time, fernandez2017compressible,ishii1999experimental,dora2014role,arakeri2004vortex,zare2008experimental,rezay2019investigation}. However, both the numerical and experimental study of the flow at this stage is inherently challenging \cite{zare2010shock,murugan2012numerical} due to the short timescales and large dynamic ranges involved. The last stage of the transient supersonic jet evolution is simply the steady underexpanded jet, which has been extensively investigated in the last decades \cite{franquet2015free}.
	
	While the structures in a transient underexpanded jet evolve in time, many of the salient flow features are analogous to those observed in the more classical steady underexpanded jet. An expansion fan originating from the nozzle exit accounts for the mismatch in pressure between the jet and the surroundings. The expansion fan reflects as compression waves from the sonic lines. These compression waves converge to an oblique shock wave, which reflects as a shock wave at the jet centerline (see illustration in figure~\ref{fig:sketch_shocklet}(b)). This reflected shock again reflects at the sonic line and results in new expansion waves. A series of reflected shock and expansion waves result in the characteristic shock cell structure or \myquote{shock diamonds} of underexpanded jets. Highly underexpanded jets are characterized by a strong Mach disk as a result of a Mach reflection at the jet centerline. As the pressure ratio decreases, the Mach disk becomes smaller. It was originally thought that for weak underexpanded jets there are no Mach reflections, but only regular reflections. However, it was shown that a regular reflection of a shock from an axis of symmetry is impossible, and therefore all reflections at the centerline of an axisymmetric jet must be Mach reflections \cite{hornung1986regular}. The flow downstream of the Mach disk becomes subsonic, while the flow downstream of the oblique shocks remains supersonic. A mismatch in temperature, entropy, and velocity occurs between the boundary of these two regions. A shear layer produced by the slipstream between the low-speed core and high speed annulus results in vortical structures, which persist across multiple shock cells \cite{edgington2014underexpanded}. In the steady state jet these aforementioned structures are well understood, however during the initial development of a transient jet their temporal evolution is far more complex, yet has received less attention.
	
\begin{figure}
		\centering
		\includegraphics[width=0.99\textwidth]{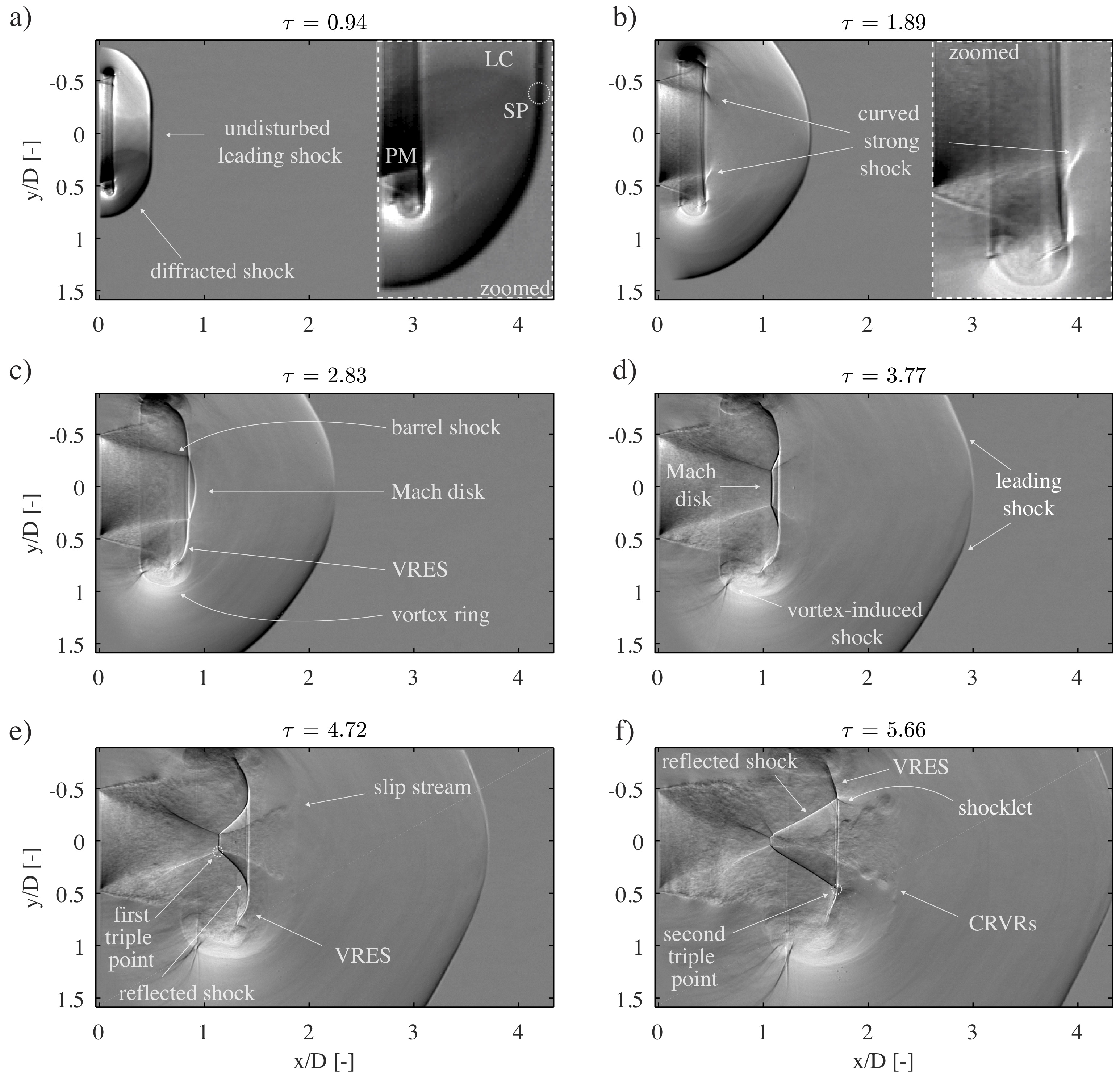}
		\caption{A time series of six \drhody schlieren images showing the detailed evolution of the exhaust flow at its early stage for \mbox{Ms = 1.76}. The x and y-coordinates are normalized by the tube diameter, \textit{D}. The origin of the axis corresponds to the point on the tube centerline at the tube exit. The time displayed above each image corresponds to the time elapsed since the shock wave has passed the pressure probe next to the exhaust tube outlet.}
		\label{fig:drho_dy_earlystage}
	\end{figure}

	To introduce the flow structures relevant to the discussion to follow,  a time series of schlieren images capturing key points in the early-stage evolution of a transient supersonic jet are presented in figure~\ref{fig:drho_dy_earlystage}, for a shock wave with a Mach number of \mbox{Ms = 1.76} exiting from a circular tube. These images are separated by a uniform time interval of 50 $\mu s$; the time $\tau$ given above the images is non-dimensionalised $\tau = (t * Ms * a_0) / D $, where \textit{t} is the time after the shock wave leaves the tube exit, \textit{D} is the tube exit diameter and $a_0$ is the speed of sound ahead of the leading shock. The corresponding experimental setup is discussed in chapter \ref{sec:Methodology}.
	
	The first stage of the jet evolution, the shock diffraction, is shown in figure \ref{fig:drho_dy_earlystage}(a). This image captures the moment immediately after the shock wave exits the tube. Towards the radial edge the shock wave has already undergone a three-dimensional diffraction as indicated by the partially curved shock; both diffracted and undisturbed leading shocks are clearly visible. The exhaust flow of the tube expands through a pseudo-steady Prandtl-Meyer expansion fan (PM) centered at the tube exit's sharp corner. The leading characteristic (LC) of these waves marks the separation point (SP) between the undisturbed leading shock and the diffracted shock. The same flow features have been observed first by \citet{skews1967shape} at a plane-walled convex corner for a diffracting shock wave. The information about the area expansion travels along the LC toward the jet centerline leading to a fully curved leading shock wave (figure \ref{fig:drho_dy_earlystage}(b)). Since the pressure at the tail of the PM expansion waves is lower than the pressure transmitted back by the leading shock wave, a second shock arises to match the two phases. \citet{friedman1961simplified} has shown that this second shock occurs due to the over-expansion caused by the increase of the volume, which does not arise in one-dimensional studies. Figure \ref{fig:drho_dy_earlystage}(b) captures the moment as this second shock is just being formed at the outer region of the jet next to the barrel shock. The upper and lower second shock waves, highlighted in figure \ref{fig:drho_dy_earlystage}(b) propagate toward the jet centerline to form a single shock wave. As these shocks coalesce, a single curved shock wave is formed (figure \ref{fig:drho_dy_earlystage}(c)). The curved shock wave transforms to a disk-shaped shock wave shortly after, as can be seen in figure  \ref{fig:drho_dy_earlystage}(d). This is the origin of the well known Mach disk of a steady underexpanded jet. 
	
\begin{figure}
		\centering
		\includegraphics[width=1\textwidth]{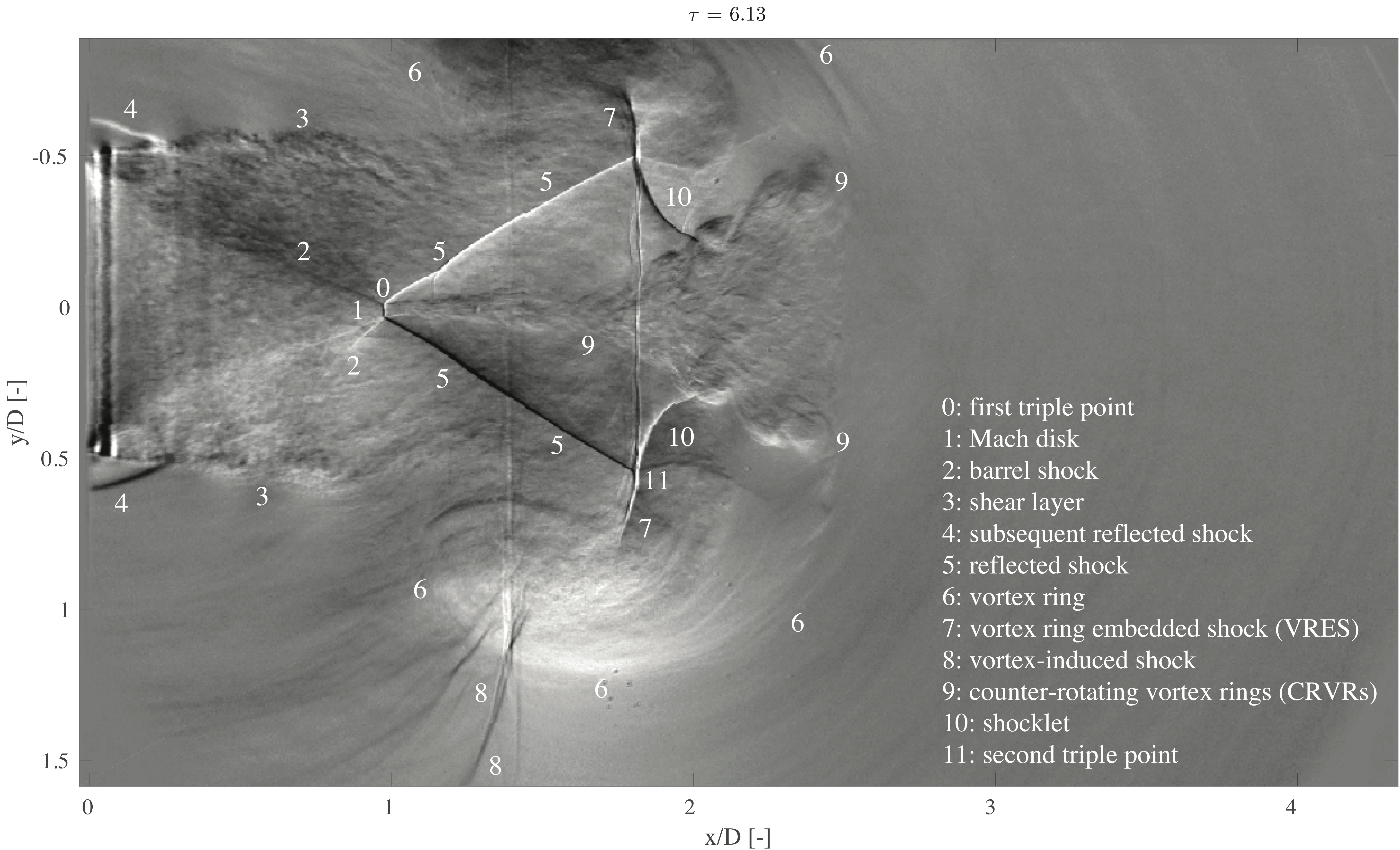}
		\caption{ Overview of the main flow features at $\tau = 6.13.$}
		\label{fig:def_x_t_and_overview}
	\end{figure}

	Besides the Mach disk, several other features are visible in figure \ref{fig:drho_dy_earlystage} that have been reported in the literature. The leading shock sets the gas inside the tube in motion by compressing the flow while propagating through the tube. Following the leading shock, a highly transient jet establishes itself at the outlet of the tube. \citet{elder1952experimental} initiated the studies of transient supersonic jets of an open-end shock tube using spark schlieren measurements. They reveal the presence of a vortex ring in the trailing jet for Mach numbers \mbox{Ms = 1.12} and 1.32,  which grows non-linearly with time and distance. In a systematic study \citet{brouillette1997propagation} found three different types of flow fields of the trailing jet depending on the leading shock Mach number. Accordingly, there is a shock-free vortex ring characterized by a very thin core for Ms \textless\ 1.43. For higher Mach numbers the vortex ring contains an embedded shock, the so-called vortex-ring-embedded shock (VRES) (also visible in figure~\ref{fig:drho_dy_earlystage}(c)). \citet{brouillette1997propagation} found the occurrence of counter-rotating vortex rings (CRVRs) for Ms \textgreater\ 1.6  (figure~\ref{fig:drho_dy_earlystage}(d)).   The primary vortex ring can also contain an additional shock wave, the so-called vortex-induced shock as indicated in figure~\ref{fig:drho_dy_earlystage}(d). These flow features can be seen more clearly in figure~\ref{fig:def_x_t_and_overview}, where an overview of the dominant flow features at $\tau = 6.13$ is given. 
	
\begin{figure}
		\centering
		\includegraphics[width=1\textwidth]{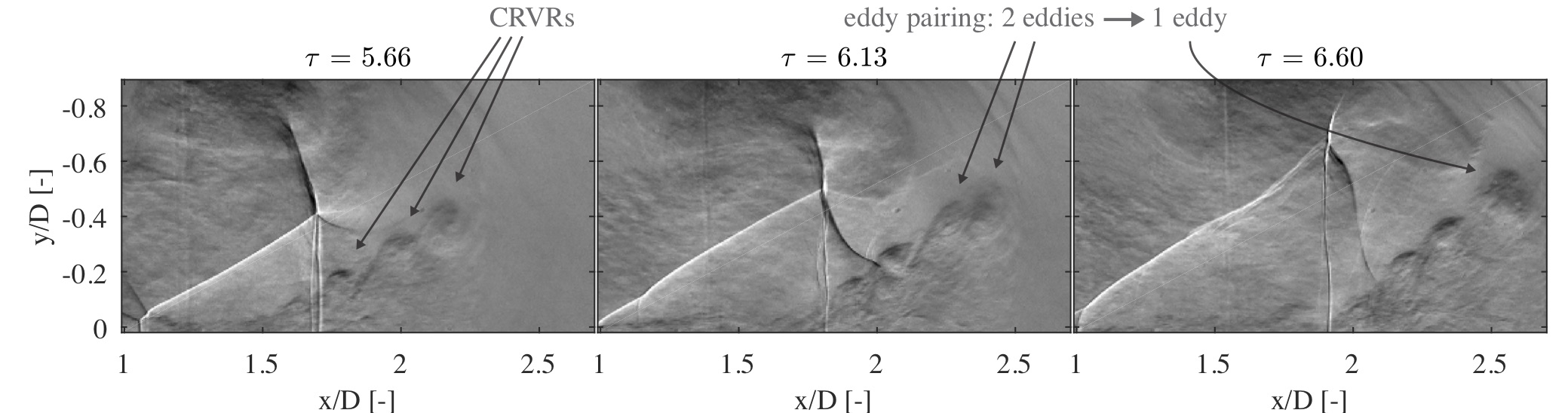}
		\caption{\drhody schlieren images showing the pairing of two eddies for \mbox{Ms = 1.76}.}
		\label{fig:eddy_pairing}
	\end{figure}

	Besides the shock waves associated with the vortex ring, the trailing jet can also contain a number of additional shock systems. \citet{ishii1999experimental} exhibited the presence of a Mach disk and a triple shock configuration in the trailing jet for high-Mach number leading shock flow. Figure~\ref{fig:drho_dy_earlystage}(d) shows the corresponding shock system, which consist of the barrel shock, the reflected shock and the Mach disk. The reflection of the barrel shock from the jet centerline as an axis of symmetry must be a Mach reflection in the same manner as for a steady underexpanded jet \cite{hornung1986regular}. The corresponding slipstream downstream of the triple point can be recognized in  figure~\ref{fig:drho_dy_earlystage}(e). Unlike the steady underexpanded jet the slipstream is inclined towards the jet boundary in radial direction.  A number of CRVRs are apparent in figure~\ref{fig:drho_dy_earlystage}(f). These vortices are generated by Kelvin-Helmholtz (KH)-instabilities of the shear layers along the slipstream~\citep{ishii1999experimental}.  \citet{dora2014role} showed that the evolution of CRVRs is driven by the same physical mechanism as for the Mach reflection. They claim that the shear layer along the slip stream grows spatially due to the eddy pairing. In accordance to that, the close-up views in figure~\ref{fig:eddy_pairing} reveal a number of eddies along the slipstream growing both in size and strength. Moreover, the image sequence shows clear evidence for the pairing process confirming the observations of \citet{dora2014role}. As shown by \citet{kleine2010timeJ}, the CRVRs wraps around the vortex ring at a later time. In a recent study, \citet{zhang2019evolution} demonstrated that the interaction of the CRVRs with the vortex ring increases the instability of the primary vortex ring. 

	The features observed in figure~\ref{fig:drho_dy_earlystage}(a)-(e) have been described in the previous studies of \citet{dora2014role}, \citet{kleine2010timeJ} and \citet{zhang2019evolution}. There is however an additional feature in figure~\ref{fig:drho_dy_earlystage}(f) that has received  far less attention: a transient shocklet formed at the intersection of the reflected shock and the VRES resulting in a second triple point. This feature has no analogue in steady-state jets, yet has received little consideration in past research on transient jets. Thus, this paper presents an experimental and numerical investigation of the shock evolution in a highly underexpanded jet. The paper is laid out as follows. A general description of the facility, as well as the schlieren methodology is given in sections \ref{sec:facility} and \ref{sec:Optical_Diagnostics}, respectively. Section \ref{sec:CFD} presents the numerical methodologies and the setup for the conducted simulations. Section \ref{sec:shocklet} considers the 
	formation and evolution of the second triple point and its associated shock structure. Section \ref{sec:mechanism} provides a proposed mechanism for the formation of the second triple point by applying a developed model for determination of post shock flow condition.  
	
	\FloatBarrier%-------------------------------------------------------------------
	\section{Methodology} \label{sec:Methodology} 
	
	\subsection{Experimental Facility and Instrumentation} \label{sec:facility}
	
\begin{figure}
		\centering
		\includegraphics[width=1\textwidth]{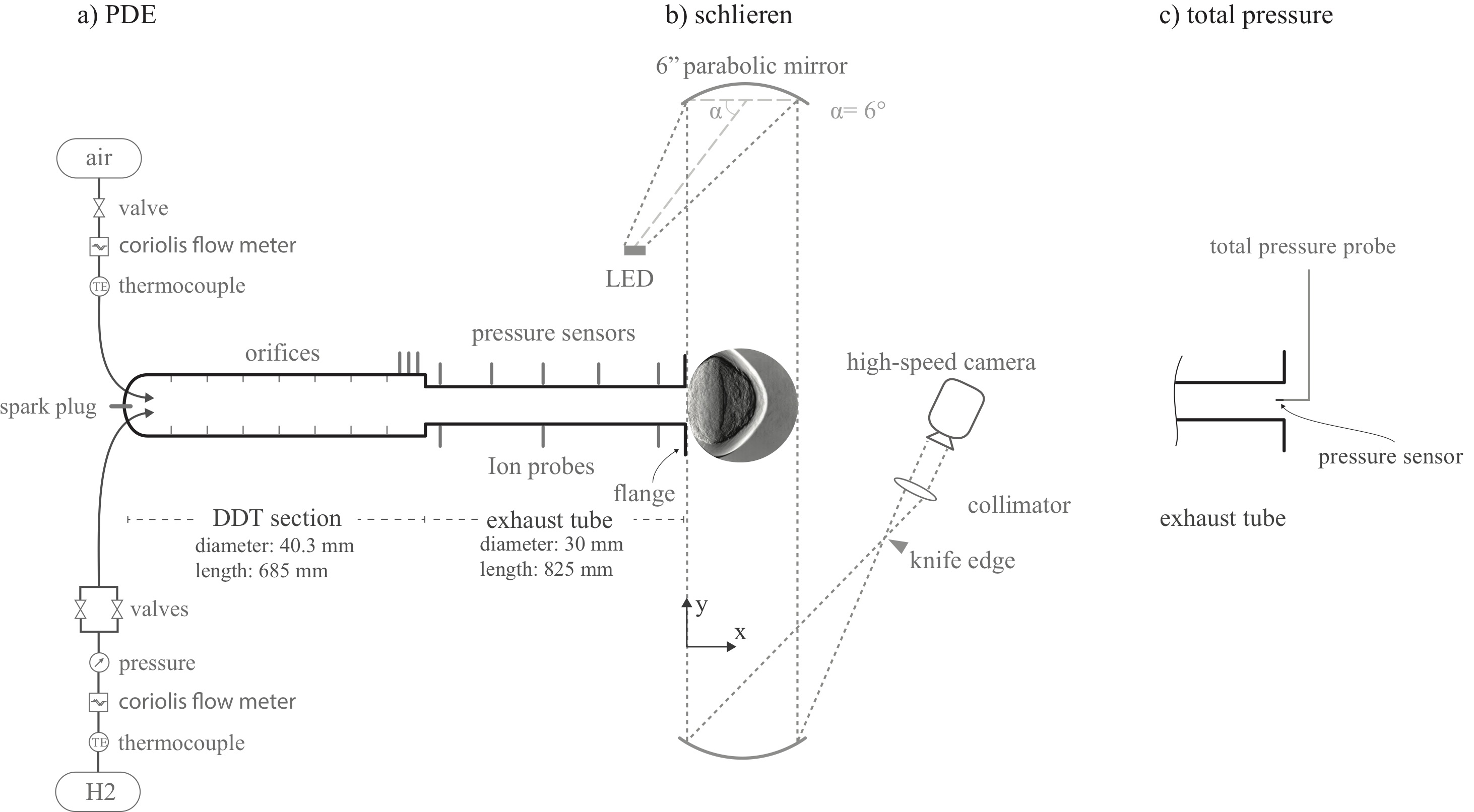}
		\caption{Sketch of the experimental setup showing (a) the pulse detonation engine, pressure sensors, and ion probes, (b) the schlieren setup and (c) the total pressure probe.}
		\label{fig:experimental_setup}
	\end{figure}

	In the current study a PDE is used to generate a shock wave. A PDE is, in its simplest form, a tube filled with a combustible mixture. Two different reactive waves can be generated using a PDE: a supersonically propagating combustion front, which is known as a detonation wave, and a subsonically propagating front, which is referred to as a deflagration wave.  A schematic of the PDE and its instrumentation is presented in figure \ref{fig:experimental_setup}. The PDE consists of two sections, the Deflagration-to-Detonation Transition (DDT) section and the exhaust tube.  Hydrogen is injected through eight circumferentially distributed fuel lines at the rear end of the tube. Air is injected directly upstream of the DDT section. Once the tube is filled with a combustible mixture a spark plug is used to initiate combustion. Orifices installed in the DDT section accelerate the flame. By varying the mixture volume and equivalence ratio the operation mode of the PDE can be adjusted. In the current study we are interested in the transient supersonic jet of a shock-induced flow and therefore, we want to minimize the impact of combustion on the exhaust flow. Hence, the tube is only partially filled with a lean mixture prior to ignition to allow for the shock wave to decouple from the reaction front. The decoupling ensures a time gap between the arrival of the shock wave and combustion products at the tube exit.
	
	The key governing parameter for the shock-induced flow is the Mach number of the shock wave propagating through the tube. To achieve comparability between the experimental and the numerical results we matched the respective Mach numbers of the leading shock waves at the tube exits, ensuring a similar flow field at the initial stage of the transient starting jet. The leading-shock Mach number for the schlieren and numerical results discussed in section \ref{sec:shocklet}  is \mbox{Ms = 2.15}.
	
    Piezoelectric pressure sensors (PCB112A05) are used to measure the leading shock wave velocity using the time of flight model. Three pressure sensors are flush-mounted in the DDT section and five in the exhaust tube. The last pressure sensor is mounted 4D upstream the tube end. The combustion front is tracked inside the tube by using ion probes flush-mounted in the opposite side to the pressure probes within the exhaust tube. These sensors are used to ensure the decoupling of the leading shock wave from the combustion front. For measurements of total pressure at the tube exit a piezoresistive Kulite XCE-062 transducer is placed at x/D = -0.3 on the jet centerline, as shown in figure~\ref{fig:experimental_setup}(c). A frequency response correction of the signal is applied by using a Kulite KSC-2 signal conditioner \cite{hurst2015real}. Two type-K thermocouples measure the temperature of air and hydrogen. The pressure in both the hydrogen and air supply lines is measured using Festo pressure transducers (SPTW-P10R). The mass flows of air and hydrogen are measured using Coriolis mass flow meters and  are controlled using proportional valves. The data from ionization and pressure probes are collected on 11 channels using a National Instruments MXI-Express DAQ system at 1 MHz sampling rate.
	
	\subsection{Schlieren Diagnostic} \label{sec:Optical_Diagnostics}	
	The flow at the open end of the PDE is investigated using time-resolved high-resolution schlieren measurements. Figure~\ref{fig:experimental_setup}(b) presents a schematic illustration of the schlieren setup.  A standard z-type configuration is used with two 6-inch parabolic f/8 mirrors for collimating and refocusing of light. A pulsed LED is used as a light source as suggested by~\citet{willert2012assessment}. A very high-intensity light pulse at a very short time span is generated using an overdriven-operated LED. An exposure time of 1 $\mu s$ has shown to be the best trade off between smearing of high-speed flow features and image contrast. The schlieren images are captured at up to 80 kHz with a Photron SA-Z camera. In the Cartesian coordinate system the x-coordinate corresponds to the jet axis and the y and z-coordinate to the radial directions. A razor blade aligned perpendicular and parallel to either the x or y-coordinate is used. The resultant images correspond to path integrated density gradients in the x-direction \drhodx and y-direction \drhody, respectively. 
	
\subsection{Numerical Simulations}\label{sec:CFD}

\subsubsection{Finite Volume Discretizations}

The present numerical simulations are based on the three dimensional Euler equations for an ideal gas, and the one-dimensional reactive Euler equations for an ideal gas mixture. In describing the respective numerical discretizations used, we will employ the following notation below: $\rho$ is the density, $\mathbf v$ the flow velocity vector, $p$ the pressure, $\mathbf Y$ the species vector, $E$ the total energy, $\mathbf I$ the identity matrix and $\gamma$ the isentropic exponent of the mixture.

The three-dimensional Euler equations read
\begin{align}\label{eq:euler}\begin{split}
   \frac{\partial}{\partial t} \rho + \nabla \cdot (\rho \mathbf{v}) &= 0\\
   \frac{\partial}{\partial t} (\rho \mathbf v) + \nabla \cdot \bigl(\rho \mathbf v \otimes \mathbf v + p \mathbf I \bigr) &= 0\\
   \frac{\partial}{\partial t} (\rho E) + \nabla \cdot \bigl(\left(\rho E + p\right) \mathbf v \bigr) &= 0
\end{split}\end{align}
and we assume the equation of states for perfect gases
\begin{align}
    \rho E = \frac{p}{\gamma - 1} + \frac 12 \rho \mathbf v \cdot \mathbf v
\end{align}
with $\gamma = 1.4$. To compute the numerical solution to \eqref{eq:euler} we use an explicit Godunov-\rk{type} second order finite volume scheme with an exact Riemann solver. The inter-cell fluxes are computed by a MUSCL reconstruction step on the conservative variables $(\rho, \rho \mathbf v, \rho E)$ and the slopes \rk{used in this} step are limited by \rk{the} van-Leer limiter  to \rk{control} artificial oscillations at discontinuities \rk{(see, e.g., \cite{Toro2009} for a textbook reference)}. Multidimensionality is handled using Strang-splitting for the spatial derivatives \rk{in} \eqref{eq:euler} and this leads in \rk{total} to a second order accurate scheme \rk{in regions of smooth solution behaviour and to first order non-oscillatory approximations near discontinuities and extrema}. \rk{This scheme is augmented by the cut-cell approach for the representation of solid wall boundary conditions introduced by Klein et~al.~\cite{Klein2009} and Gokhale et al.~\cite{Gokhale2018} which is compatible with directional operator splitting.} The cylindrical boundary of the combustion tube is represented as a level set and is embedded in a regular Cartesian grid. We also make use of block-structured adaptive mesh refinement techniques (Berger \& Oliger~\cite{BergerOliger1984}) to locally refine the grid in regions of interest, such as shock waves or cut-cells.

\rk{The one-dimensional reactive gas flow simulations in the detonation tube are based on the one-dimensional form of \eqref{eq:euler} with chemical reactions described by balance laws for the chemical species, 
\begin{align}
   \frac{\partial}{\partial t} (\rho \mathbf Y) + \nabla \cdot \bigl( \rho u \mathbf Y \bigr) &= - \rho \dot {\mathbf Y}\,.
\end{align}
Furthermore, the energy equation of state is modified to account for a mixture of gases, i.e., 
\begin{align}
	\rho E = \rho \int_{T_0}^{T} c_v(\tau, \mathbf Y)\, \mathrm{d} \tau + \rho Q_0(\mathbf Y) + \frac 12 \rho u^2\,,
\end{align}
where $c_v(\tau, \mathbf Y)$ is the specific heat capacity at constant value and the formation enthalpy at $T = T_0$ is $Q_0(\mathbf Y)$. In this study these functions, just as the reaction rate functions $\dot {\mathbf Y}(T,p,{\mathbf Y})$ are provided by an H2-O2 reaction mechanism for high pressure combustion following Burke et al.~\cite{Burke2011}.}

\rk{The numerical scheme used here is described in \cite{BerndtEtAl2016}. It differs from the inert gas 3D solver explained above by (i) the use of Strang splitting for the implementation of the chemical reaction terms, and (ii) the use of the HLLE approximate Riemann solver as a numerical flux function. The HLLE solver is the version of the general HLL scheme of Harten et al.~\cite{HartenEtAl1983} with the numerical signal speeds determined according to Einfeldt~\cite{Einfeldt1988}. This flux function provides added robustness and efficiency relative to the exact Riemann solver. Its advantages for detonation wave applications have been discussed by Berndt~\cite{Berndt2014}.}

\subsubsection{Initial Data and Boundary Conditions \rk{for the Approximate Representation of the Experiment}}

\begin{figure}
            		\centering
                \includegraphics[width=0.65\textwidth]{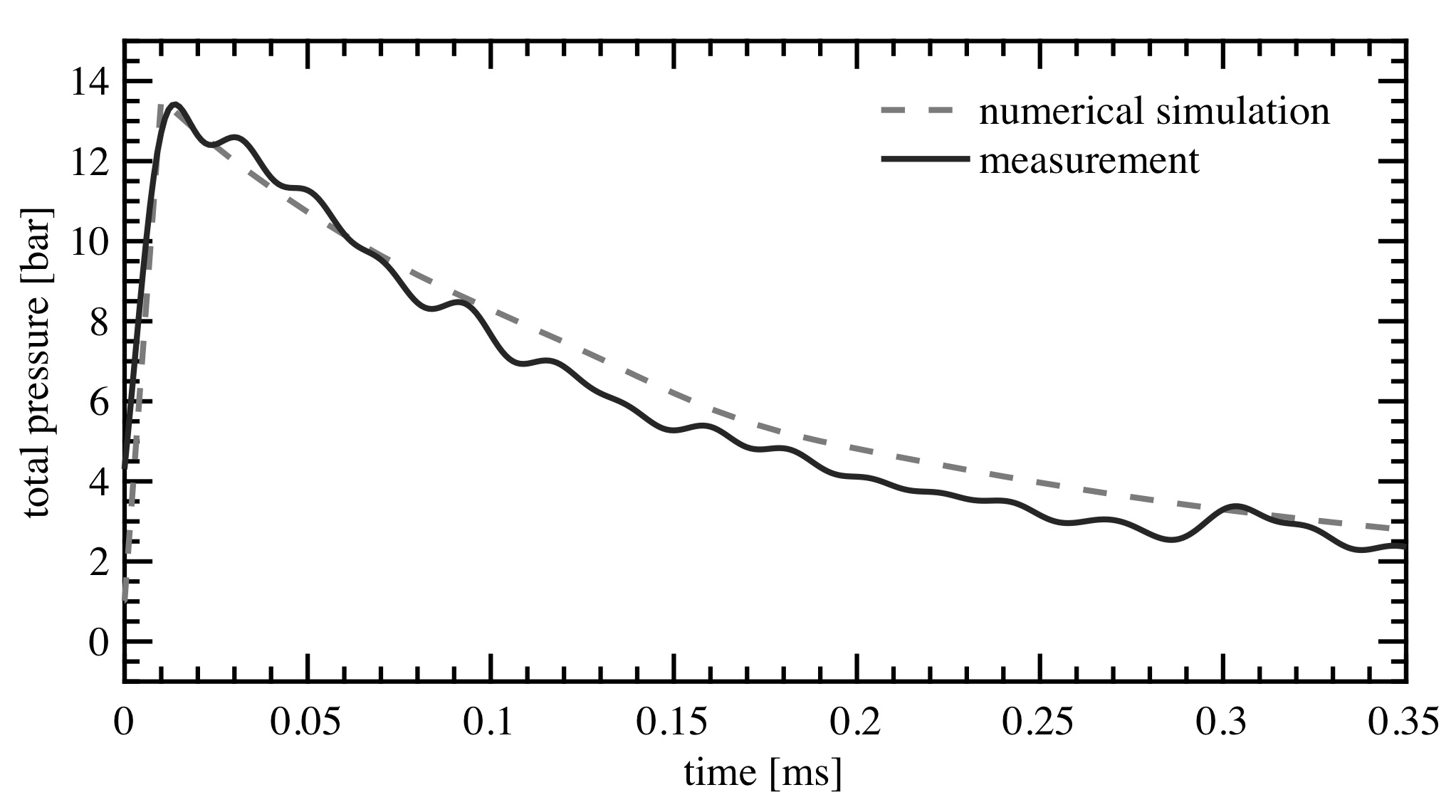}
            		\caption{Total pressure over time for \mbox{Ms = 2.15} at \mbox{x/D = -0.3}.}
            		\label{fig:pressure_cfd_vs_exp}
        	\end{figure}

The combustion tube in the experiment is only partially filled with the combustible mixture but its fill fraction and equivalence ratio along the tube is not known. Although we are given \rk{measured} pressure data over time at the tube outlet in addition to the \rk{M}ach number for the leading shock wave, a complete description of the thermodynamic quantities \rk{is not \rk{experimentally} available}. To approximate inflow boundary conditions \rk{into the 3D simulation domain} at the outlet \rk{of the combustion tube,} we perform a series of one-dimensional simulations of H2-O2 detonations \rk{and compared the results with the measurements. Varying the equivalence ratio of the combustible mixture and the fill fraction of the tube we found a one-dimensional solution within the tube that matches the Mach number of the leading shock as well as the total pressure over time (figure \ref{fig:pressure_cfd_vs_exp}) rather accurately}.

\rk{Even if the flow states in the combustion tube can be well approximated by cross-sectional averages of the conserved quantities, thereby allowing for a one-dimensional approximation, the flow states next to the tube exit are always affected non-trivially by multidimensional effects. To properly capture these, the three-dimensional simulations cover the entire length of the PDE in addition to a large flow domain beyond the tube exit. Initial conditions within the tube are given by the solution from the one-dimensional computation at a point in time right after the combustible mixture is entirely consumed, but before the leading shock wave has reached the tube exit. Outside the combustion tube we initially assume air at rest at atmospheric conditions.} 

\FloatBarrier%-------------------------------------------------------------------
\section{The early-stage evolution of the transient supersonic jet}
In the following the formation and evolution of the second triple point and its associated shock structure are discussed based on both experimental and numerical results. Finally a proposed mechanism for the second triple point is presented. 
    
\subsection{Formation and evolution of the shocklet }\label{sec:shocklet}
In figure 2 we noted the presence of a second triple point at the intersection of the reflected shock and the vortex ring. Unlike the first triple point and its associated system of shocks, this second triple point is not observed in a steady underexpanded jet. While the shocks associated with this second triple point are visible in some published work, as of yet there has been no discussion of the mechanism by which it forms. The triple point is formed at the intersection of the reflected shock, the VRES, and a new transient shock structure which, due to its transient nature, we will refer to as a shocklet. The shocklet and its associated triple point are only present for a short time during the early evolution of the transient jet. 
	
\begin{figure}
		\centering
		\includegraphics[width=0.7\textwidth]{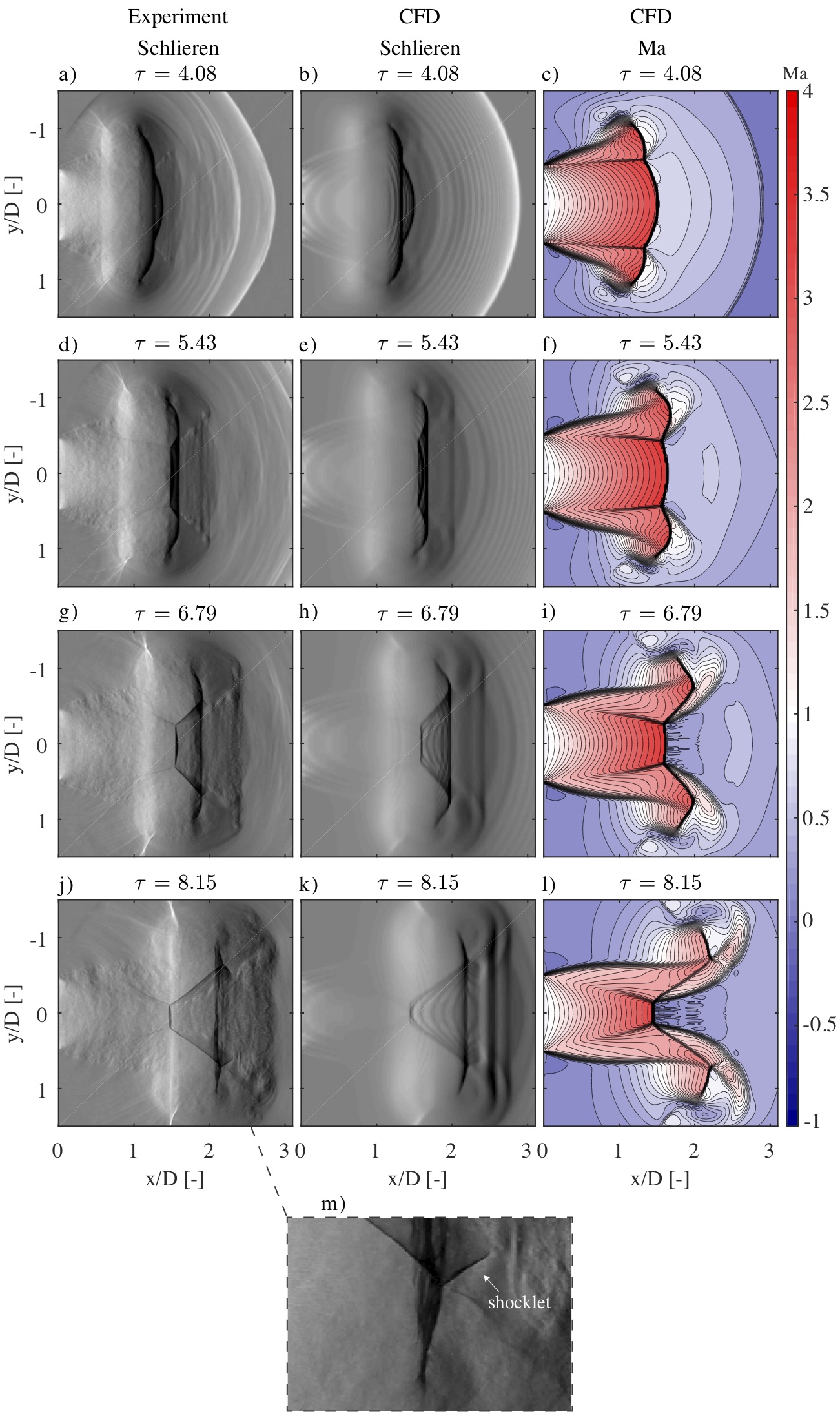}
			\caption{Formation of the shocklet based on experimental \drhodx schlieren images for \mbox{Ms = 2.15} (left column) numerical \drhodx  schlieren images (center), and Mach number contour plots at z = 0 (right column).}
		\label{fig:drho_dx_earlystage}
	\end{figure}

Figure~\ref{fig:drho_dx_earlystage} presents a series of numerical and experimental snapshots spanning this early evolution period. Experimental schlieren images are compared with numerical schlieren images produced by path integration through the three-dimensional simulation data. The numerical schlieren images display grey scales of the quantity $$ S(x,y) = \int_{z_0}^{z_1}\frac{\partial \rho}{\partial x}(x,y,z) dz.$$ In addition, a planar representation of the early evolution of the underexpanded jet is given  in figure~\ref{fig:drho_dx_earlystage}  by  contour plots of the Mach number at z = 0. Subsonic and supersonic region of the jet cross section are color-coded with blue and red colors, respectively. There is excellent agreement between the numerical and experimental schlieren results during this early evolution of the jet. Figure~\ref{fig:drho_dx_earlystage}(a)-(c) shows the leading shock wave and a triple point configuration consisting of the barrel shock, the Mach disk and the reflected shock. Up to this time ($\tau \leq 4.08$) the reflected shock and the VRES are simply the same shock wave, and the size and position of the shock are primarily dictated by the strong vortex in which it is embedded. As the vortex propagates further downstream, the upstream boundary condition for the shock is instead dictated by the Mach reflection arising from the incident shocks generated at the lip. At larger radial positions however, the shock is still very much a function of the velocity field induced by the vortex ring. At $\tau = 5.43$, while the shock still forms a contiguous surface, the angle set by the triple point is significantly different to that required in the vortex ring. This disparity increases as the vortex ring propagates further from the nozzle, and while the shock surface remains contiguous, by $\tau = 6.79$ a sharp kink forms on this surface, separating the the VRES from the reflected shock (figure~\ref{fig:drho_dx_earlystage}(g)-(i)). As the vortex ring moves further downstream at $\tau = 8.15$, the kink becomes a triple point, and a second triple shock configuration occurs (figures~\ref{fig:drho_dx_earlystage}(j)-(m)). The shocklet is clearly visible between the reflected shock and the CRVRs in both experimental and numerical data as shown in figures~\ref{fig:drho_dx_earlystage}(j)-(m).
	
\begin{figure}
		\centering
		\includegraphics[width=1\textwidth]{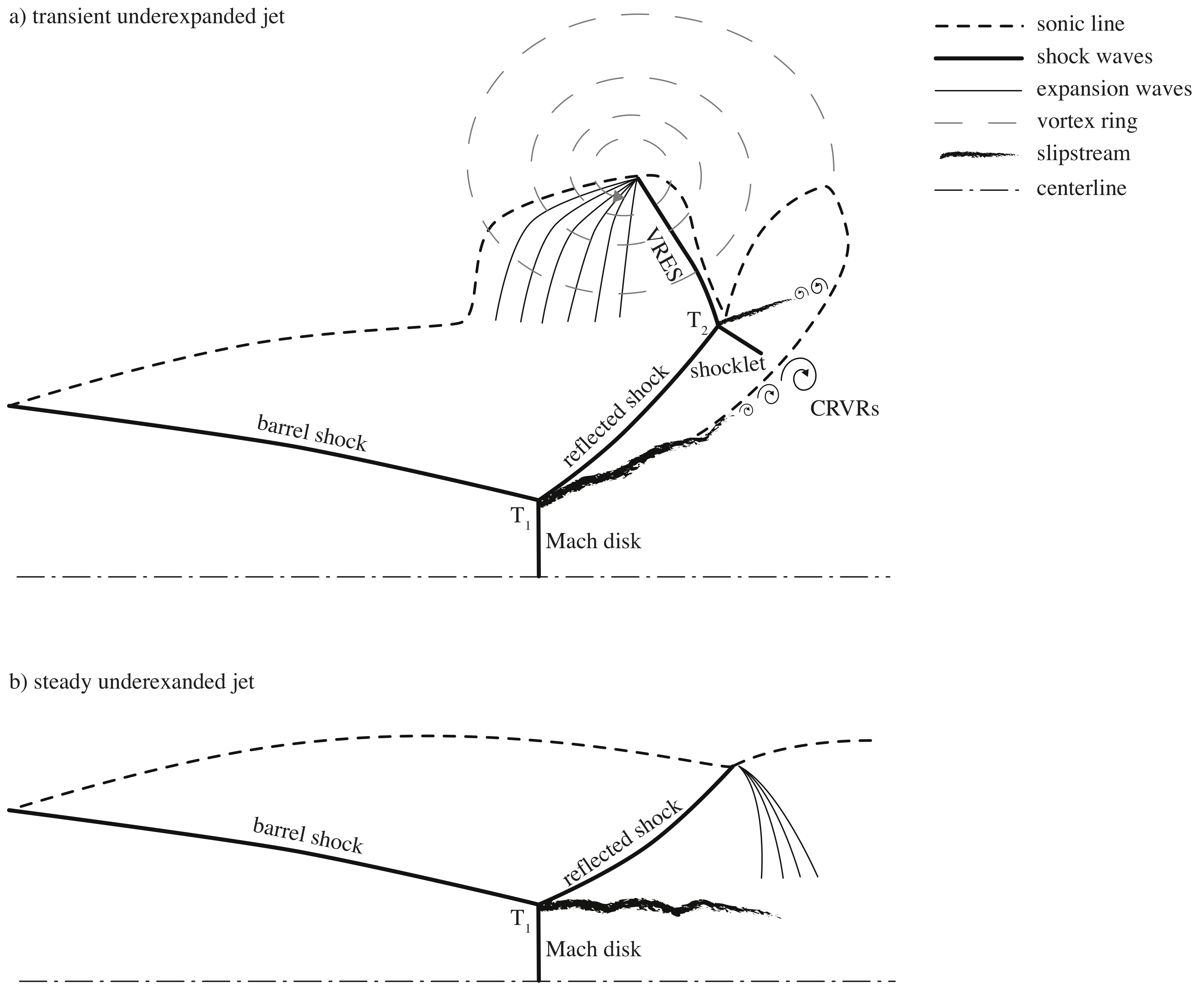}
		\caption{Schematic illustration of the jet structure for the transient and steady underexpanded jet.}
		\label{fig:sketch_shocklet}
	\end{figure}

To facilitate a clearer description of the formation of the shocklet, figure~\ref{fig:sketch_shocklet} presents a schematic comparison between the structures in the transient jet and those in its steady-state counterpart. The illustration in  figure~\ref{fig:sketch_shocklet}(a) corresponds approximately to the flow state shown in figures~\ref{fig:drho_dx_earlystage}(j)-(m). A triple point configuration as a result of a Mach reflection can be observed for both steady and transient jet. This shock system consists of the barrel shock, the Mach disk, the reflected shock and the triple point (T1). In case of the steady underexpanded jet, the reflected shock of the primary shock system reflects as expansion waves from the sonic line (figure~\ref{fig:sketch_shocklet}(b)). However, for the transient jet the sonic line is significantly distorted by the presence of the vortex ring, and a simple reflection does not occur. Instead, a second triple shock configuration occurs, as illustrated in figure~\ref{fig:sketch_shocklet}(a). This shock system consists of the reflected shock, the VRES and the shocklet, intersecting at a second triple point T2. 
	
	\begin{figure}
		\centering
		\includegraphics[width=0.9\textwidth]{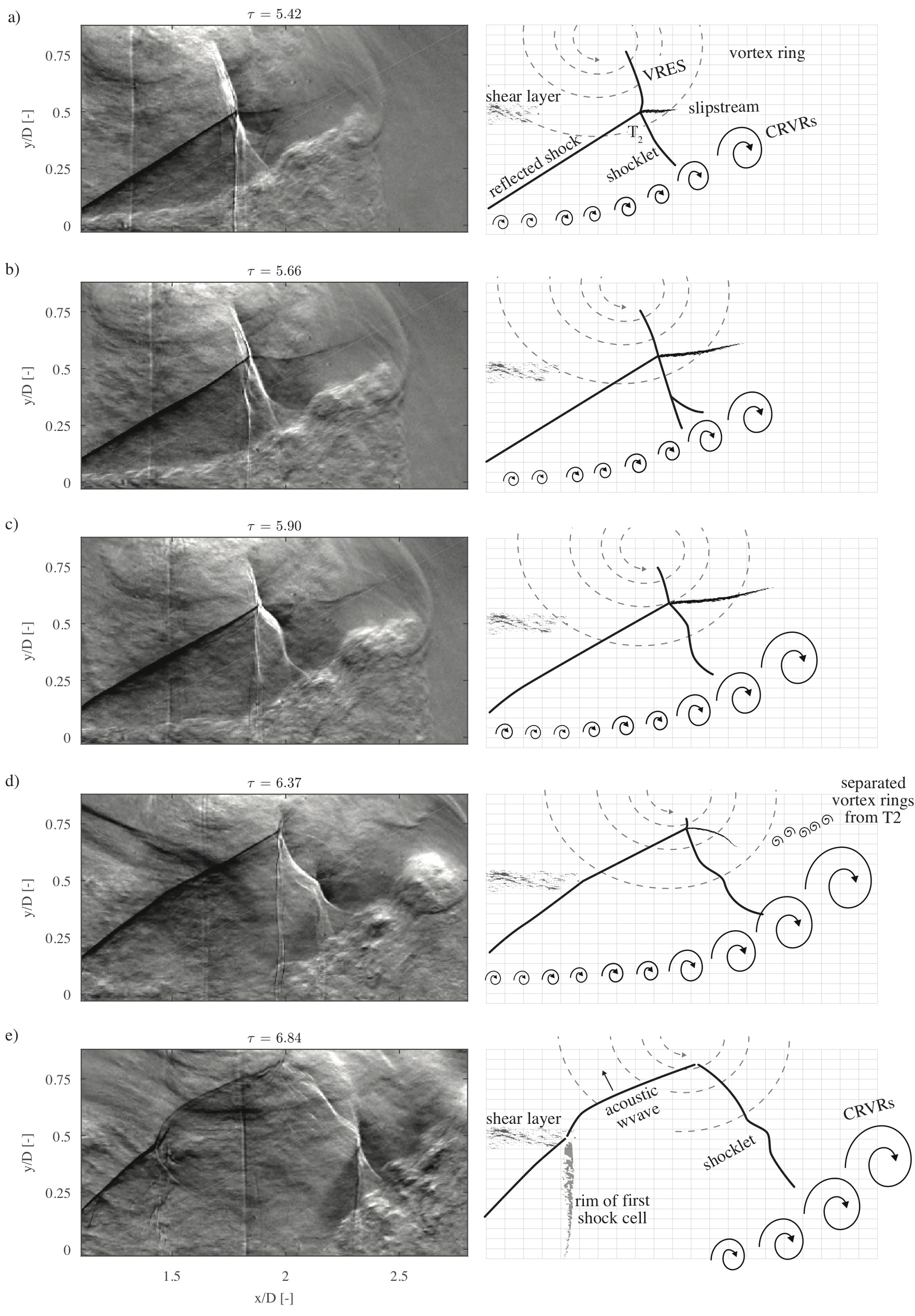}
		\caption{\drhody schlieren images for \mbox{Ms = 1.76} showing the evolution of the second triple point configuration.}
		\label{fig:shockletEvolution}
	\end{figure}
	 
The subsequent evolution of the second shock system is shown in Figure~\ref{fig:shockletEvolution} based on a time series of five schlieren images. No further tilting of the reflected shock toward the jet center line can be observed; the initial angle of the reflected shock has reached an approximate steady state. In contrast, the vortex ring and its embedded shock (VRES) move further downstream, with the VRES decreasing continuously in size. Therefore, the triple point T2 translates downstream and radially outwards  (figure~\ref{fig:shockletEvolution}(a)-(e)). As the vortex ring convects further, the sonic line shifts inwards, and a portion of the reflected shock must become propagative; this upstream-propagating wave rapidly decays into an acoustic wave, in a manner analogous to the shock leakage process of jet screech \cite{edgington2019aeroacoustic}. The conversion of this part of the reflected shock into an upstream-propagating wave effectively terminates the second triple point (figure~\ref{fig:shockletEvolution}(e)). While the second triple shock configuration and its corresponding slipline terminate, small vortex rings along the slipline separate from the triple point (T2), as shown in  figure~\ref{fig:shockletEvolution}(d). Finally, the impingement point of the reflected shock upon the jet shear layer appears as a wavy line in the schlieren image, representing the rim of the first shock cell. Also evident in figure~\ref{fig:shockletEvolution} is an interaction between the primary and second triple point configuration. The upper bound of the shocklet is the triple point, its lower bound is the sonic line associated with the internal shear layer generated by the first triple point. The shocklet undergoes deformation via interaction with the CRVRs generated along the primary slipline, (Figure~\ref{fig:shockletEvolution}(b)-(d)) and the interaction produces lambda-shocks close to the CRVRs (figures~\ref{fig:shockletEvolution}(b)). Both, the experimental and numerical data show a second triple point configuration. 
   
\subsection{A proposed mechanism for the formation of the second triple point}\label{sec:mechanism}

\begin{figure}
	\centering
	\includegraphics[width=1\textwidth]{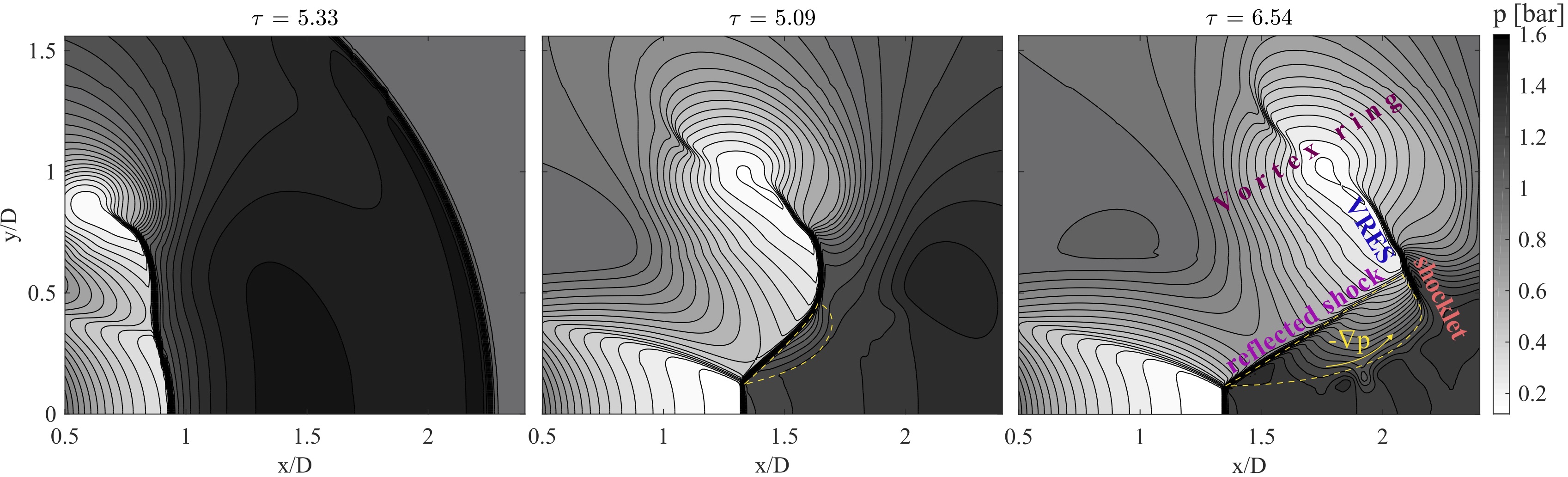}
	\caption{Sequence of pressure contours determined from numerical simulations for a Riemann problem with \mbox{Ms = 1.71}. The yellow dashed line indicate the formation of a pressure gradient that potentially leads to the formation of the shocklet.}
	\label{fig:pressure_shocklet_overview}
\end{figure}

The formation of the second triple point in the transient jet has no equivalent in the steady jet, as visualized in figure~\ref{fig:sketch_shocklet}. Thus, the explanation for its formation must lie inherently in the dynamics of a transient jet. The convection of the vortex ring and its associated shock structure is one such process, and the temporal variation in upstream flow conditions within the tube is another. In order to separate these processes, the numerical simulations were repeated with a constant inflow condition at the tube exit. Also, a different shock Mach number of \mbox{Ms = 1.71} is chosen, to rule out the impact of the shock strength. The inflow conditions for the numerical simulations corresponds to the solution of the Riemann problem for a planar shock wave propagating at a constant speed corresponding to \mbox{Ms = 1.71}. The results are shown in figure~\ref{fig:pressure_shocklet_overview}, where the formation of the second triple point is clearly visible. Hence, this suggests that the transient interaction between the first triple point and the vortex ring are the likely explanation for the formation of the second triple point.

The consideration of the time-series of pressure distributions shown in figure~\ref{fig:pressure_shocklet_overview} demonstrates that unlike the steady jet counterpart, the reflected shock is non-stationary for a time before the formation of the second triple point. The reflected shock elongates and rotates, as the vortex ring and its embedded shock move further downstream. As the reflected shock tilts toward the jet centerline, a strong pressure gradient parallel to the shock develops along its downstream face. This region of negative pressure gradient, from the jet core to the jet shear layer, is marked as \NPG in figure~\ref{fig:pressure_shocklet_overview}. Unlike for a steady jet, this region of negative pressure gradient grows in size and strength with time for a transient jet.  We suggest that it is the motion of the reflected shock, which results in a pressure gradient downstream of the reflected shock, that in turn leads to the formation of the shocklet and the second triple point. In the following we develop a model to evaluate this hypothesis.

	\subsubsection{Transient oblique shock model (TOS)}\label{sec:TOS}
	
	To test the hypothesis that it is the unsteady motion of the reflected shock that gives rise to the formation of the triple point, we develop a model for the effect of this motion. The model delivers the post-shock flow conditions for a transient, rotating oblique shock wave based on the pre-shock flow conditions and the shock motion. The underlying assumption of the approach is that the moving shock wave can be treated as a quasi-steady problem by converting the flow velocity into a reference frame that moves with the shock.
	
\begin{figure}
	 		\centering
	 		\includegraphics[width=0.65\textwidth]{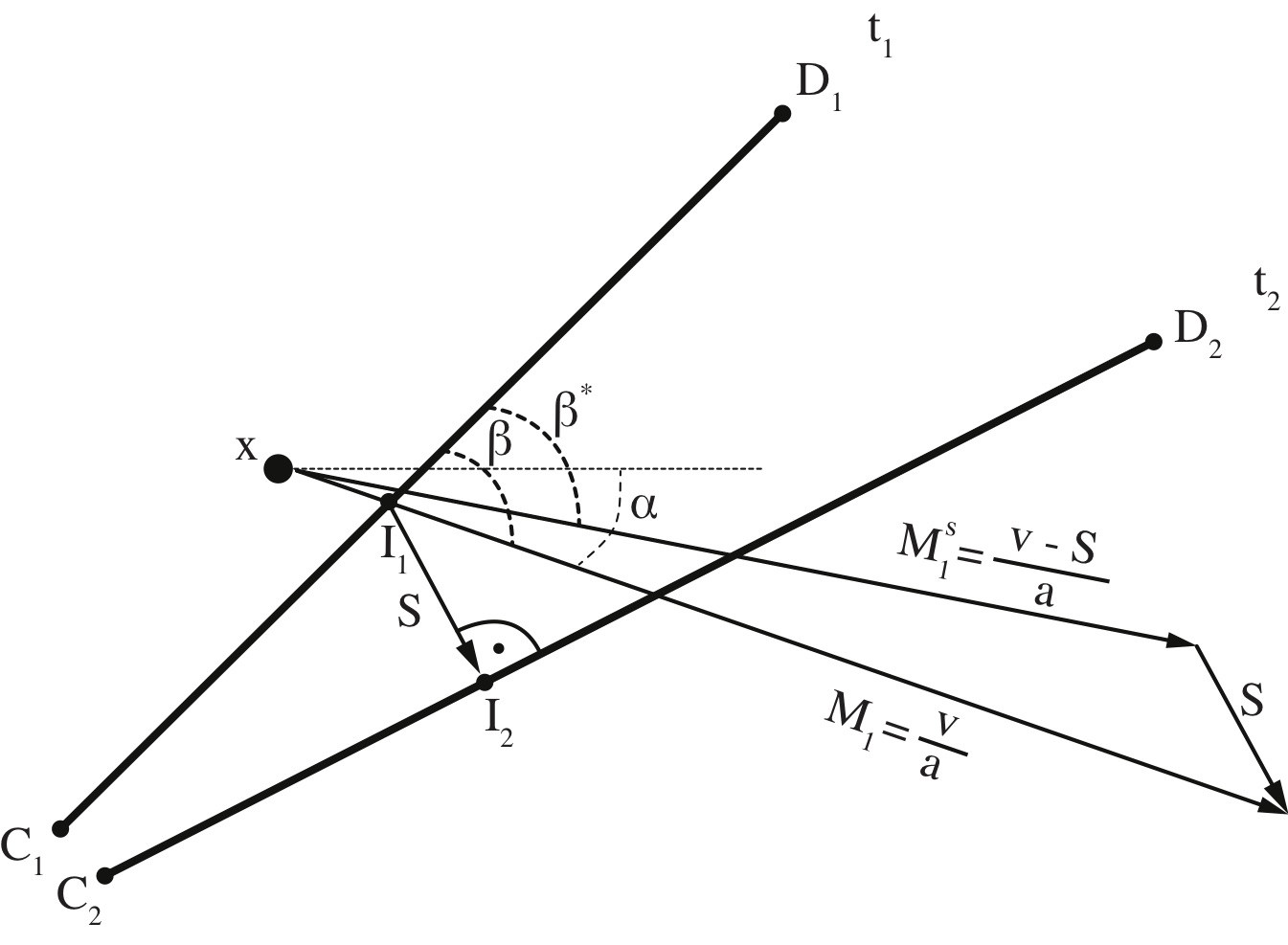}
	 		\caption{Schematic illustration of the transient oblique shock (TOS) model}
	 		\label{fig:prep_foot_sketch}
	 	\end{figure}

	 A schematic illustration of the problem is presented in figure~\ref{fig:prep_foot_sketch}(a). In a time period of $\Delta t = t_2 - t_1 $ a shock wave moves from $\overline{C_1D_1}$ to  $\overline{C_2D_2}$. The objective of the model is to determine the post-shock condition for a particle upstream of the shock wave at $x$, which will be processed by the shock wave after a certain time. The particle in $x$ at $t_1$ has already passed the shock wave at $t_2$, since the flow velocity $v$ is higher than the corresponding shock velocity $S$. 
	 
	A simple approach is used to estimate the shock velocity $S$.  The intersection of $v$ with $\overline{C_1D_1}$ is marked as a point $I_1$ in figure~\ref{fig:prep_foot_sketch}.  A perpendicular line from $I_1$ to $\overline{C_2D_2}$ intersect with $\overline{C_2D_2}$  at a point $I_2$. The shock velocity $S(x)$ is approximated simply by the displacement of the shock $\overline{I_1I_2}$ over the time interval by $S =  \frac{\overline{I_1I_2}}{\Delta t}$.

	The determination of post-shock properties for an oblique shock is an elementary gas-dynamics problem, solved by the simple application of the Rankine-Hugoniot equations. The problem here however involves a shock that is both translating and rotating. The proposed model is thus essentially an attempt to produce an appropriate co-ordinate transformation to allow the application of quasi-steady 1-D conservation equations to a rotating shock. Therefore, the Mach number in the absolute reference $M_1$ = $\frac{v}{a}$ must be converted into a reference frame that moves with the shock. Taking the shock velocity into account, the Mach number in the shock reference frame is simply $M^{\scales}_1 = \frac{v-S}{a}$. To apply the Rankine-Hugoniot equations for an oblique shock, the normal component of $M^{\scales}_1$ is determined by considering the shock angle in the shock reference frame $\beta^{\scales}$. As shown in figure~\ref{fig:prep_foot_sketch}, $\beta^*$ is the cross angle between $\overline{C_1D_1}$ and $ M^{\scales}_1$. Assuming an infinitesimal $|\overline{I_1I_2}|$, the mean value of $\beta$ and $\beta^*$ is taken as the shock angle $\beta^{\scales}$. Hence, the normal Mach number in shock reference $M\text{n}^{\scales}_1$  can be determined as $M\text{n}^{\scales}_1 = M^{\scales}_1 \sin \beta^{\scales}$. Finally, the post-shock conditions are evaluated by applying the normal component of the Mach number in the Rankine-Hugoniot equations:
	
	\begin{align}
	M_2^2 = \frac{1 + \frac{\gamma - 1}{2} (M^{\scales}_1 \sin \beta^{\scales})^2}{\gamma (M^{\scales}_1 \sin \beta^{\scales})^2 - \frac{\gamma - 1}{2}},
	\label{eqn:oblique_shock_transient}
	\end{align}
	
	\begin{align}
	\frac{p_2}{p_1} = 1 + \frac{2\gamma}{\gamma + 1} ((M^{\scales}_1 \sin \beta^{\scales})^2 - 1),
	\label{eqn:p_transient}
	\end{align}
	
	\begin{align}
	\frac{\rho_2}{\rho_1} = \frac{(\gamma + 1) (M^{\scales}_1 \sin \beta^{\scales})^2}{1 + (\gamma-1) (M^{\scales}_1 \sin \beta^{\scales})^2},
	\label{eqn:rho_transient}
	\end{align}
	
	\begin{align}
	\frac{T_2}{T_1} = \frac{p_2}{p_1} \frac{\rho_1}{\rho_2}.
	\label{eqn:T_transient}
	\end{align}
	Here, $p$, $\rho$, $T$ are the pressure, density, and temperature, respectively. In the following section, this model is used to demonstrate the formation mechanism of the second triple point. 
	
	\subsubsection{Formation mechanism of the triple point based on the TOS model}\label{sec:formation_machanism}

 As previously stated, the formation of the secondary triple point must be linked to the transient evolution of the jet, as only the first triple point appears in steady-state underexpanded jets. The secondary triple point is made up of the reflected shock from the first triple point, the VRES, and the shocklet, as shown in figure~\ref{fig:sketch_shocklet}(a). Of these, the first two have been discussed in the literature at some length; the shocklet is the component that has therefore gone undescribed. To establish why the shocklet forms, we apply the TOS model to the motion of the reflected shock, with a starting point well before the shocklet is observed. Initial observations suggest that the pressure gradient parallel to the downstream face of the reflected shock is likely linked to the shocklet's formation, thus  we seek to test whether this pressure gradient is a result of the motion of the shock. 

	\begin{figure}
		\centering
		\includegraphics[width=1\textwidth]{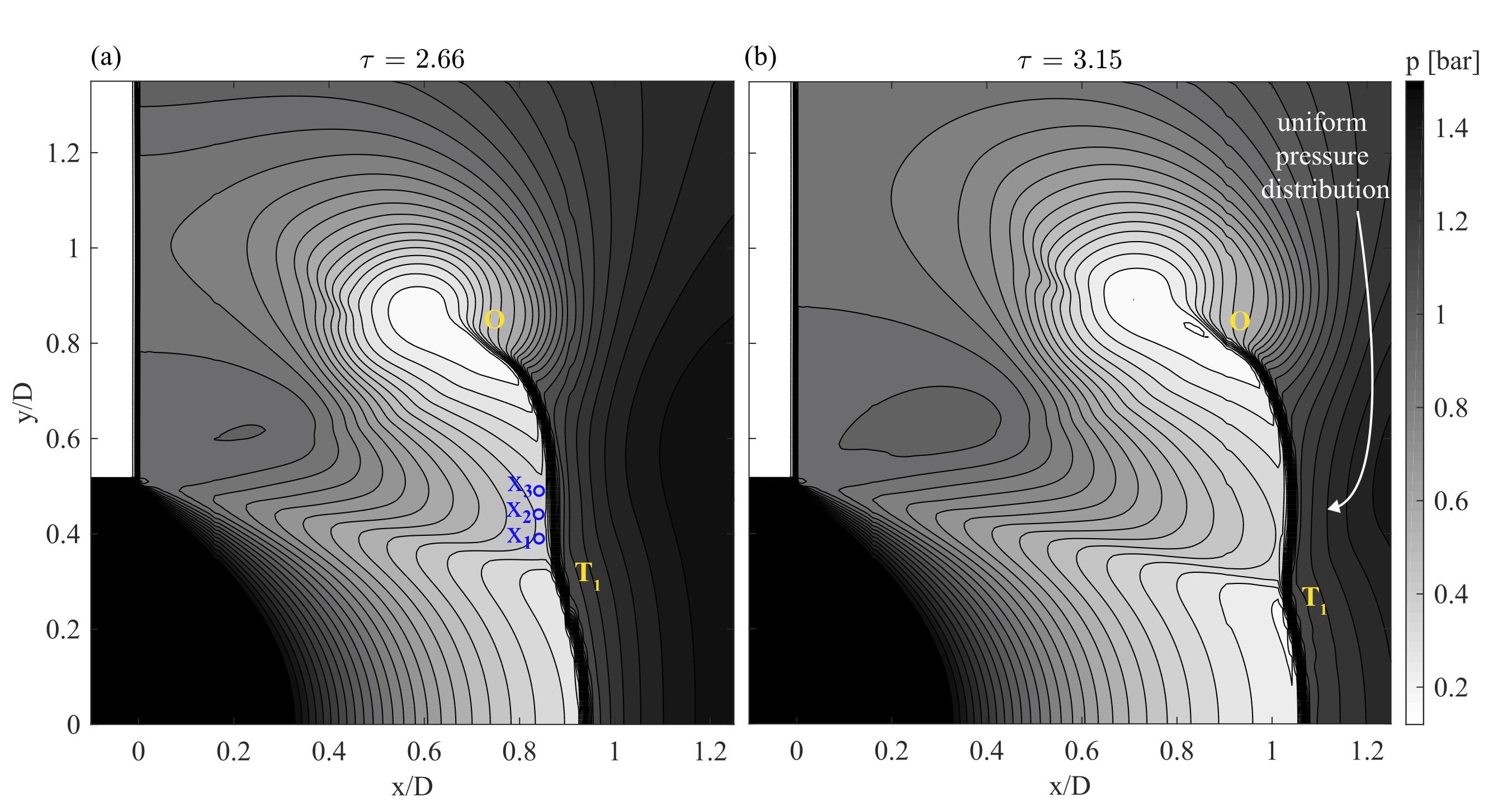}
		\caption{(a) Pressure distribution at \tZwanzig as derived from the numerical simulations at constant inflow conditions with \mbox{Ms = 1.71}. $\text{T}_1$ and O represents the first triple point and the tail of the VRES.  The TOS model is applied for $x_1$, $x_2$ and $x_3$. The result of the TOS analysis is given in table \ref{tab:p1_k20_p_an_uniform}. (b) Pressure distribution at \tZweiUndZwanzig shows a uniform pressure distribution downstream of the reflected shock.}
		\label{fig:k20_p_an_uniform1}
	\end{figure}
	
	 The application of the TOS model involves analysis of a series of discrete points parallel to the upstream face of the reflected shock ($x_1$ to $x_n$), as per  figure~\ref{fig:k20_p_an_uniform1}. We use the early stages of evolution before the formation of the shocklet to test the validity of the model. Thus we start our analysis at \tZwanzigY, where the reflected shock and VRES are essentially a single contiguous shock wave (from $\text{T}_1$ to O in figure~\ref{fig:k20_p_an_uniform1}(a)). Three discrete points $x_1$ to $x_3$ are selected just upstream of the reflected shock for the TOS analysis. The changes in flow properties during passage through the shock wave are considered for particles originating at these points using the TOS model. To this end, the motion of the reflected shock is tracked for two snapshots (\tZwanzig and \tEinundZwanzigY).  A small time period of $\Delta \tau = 0.24$ is chosen for the TOS analysis, as an infinitesimal shock displacement is the underlying assumption of the model. The results of the TOS analysis are presented as several input and output parameters in table \ref{tab:p1_k20_p_an_uniform}. Here the input parameters, $M_1$, $p_1$  , $\alpha$ are the Mach number, pressure and flow angle, respectively, extracted directly from the numerical results. The output parameters of the TOS analysis, $M\text{n}^{\scales}_1$, $M^{\scales}_1$, $\beta^{\scales} $ and $p_2$ are the normal Mach number in shock reference, the Mach number in shock reference, the shock angle and the pressure downstream of the reflected shock, respectively. 

% k 20 - case 15 keynote
	\begin{table}
		\begin{small}
			\caption{TOS results for \tZwanzig shown in figure~\ref{fig:k20_p_an_uniform1}(a). \label{tab:p1_k20_p_an_uniform}}
			\begin{tabular}{lccccccccc} 
				&\multicolumn{3}{c}{input} &&\multicolumn{4}{c}{output} \\
				\cline{2-4}
				\cline{6-10}
				                  &$M_1$  &      $p_1$  &          $\alpha$           & & $M\text{n}^{\scales}_1$&      $M^{\scales}_1$   & $\beta^{\scales} $  & $S [m/s]$  & $p_2 [bar]$ \\ 
				\midrule
				{$\boldsymbol x_1$}		& 2.22 &    0.43    &          9.7      &&        1.60     &           1.62     & 79.9     & 188&  \textbf{1.2}    \\     
				{$\boldsymbol x_2$}     & 2.24 &    0.43   &         10.8    &&         1.60        &            1.64    &   78.5    & 188 & \textbf{1.2}    \\     
				{$\boldsymbol x_3$}     & 2.28  &   0.4    &         12.5    &&        1.63      &            1.68    &  76.2        &  188 & \textbf{1.2}    \\     
				\\                 
			\end{tabular}
		\end{small}
	\end{table}

Figure~\ref{fig:k20_p_an_uniform1}(b) presents the pressure distribution based on the numerical simulation, shortly after the distribution in  figure~\ref{fig:k20_p_an_uniform1}(a). This time interval $\Delta \tau = 0.48$ allows for the particle upstream of the reflected shock at \tZwanzig to be processed by the shock wave at \tZweiUndZwanzigY. For the TOS analysis the calculated pressure downstream of the shock wave is given  as $p_2$ in table~\ref{tab:p1_k20_p_an_uniform}. The results show a  constant value of  1.2 bar for $x_1$, $x_2$ and $x_3$, i.e., a uniform pressure distribution. In accordance, the results from the numerical simulations confirm a uniform pressure distribution downstream of the reflected shock in figure~\ref{fig:k20_p_an_uniform1}(b). Hence, the predicted uniform pressure distribution downstream of the reflected shock based on the TOS analysis agrees qualitatively very well with the CFD results. This agreement sustains for the entire conducted analysis, as pointed out in the reminder of this section. Hence, the TOS model is considered as valid. 
	
\begin{figure}
		\centering
		\includegraphics[width=0.9\textwidth]{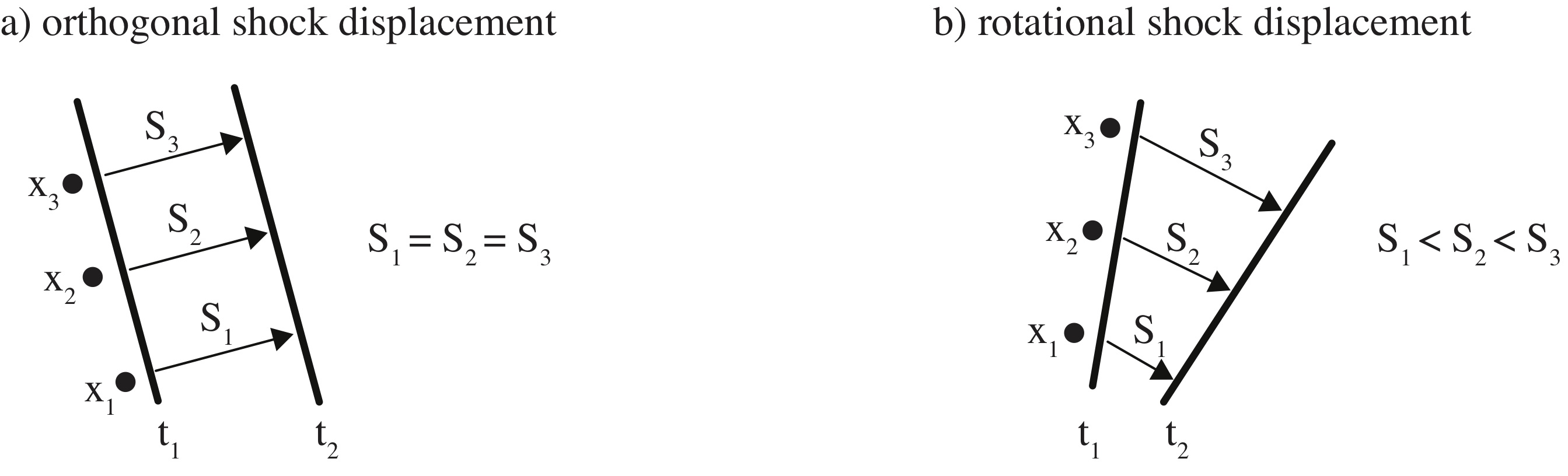}
		\caption{Shock propagation velocity for (a) orthogonal and (b) rotational shock displacement.}
		\label{fig:illustration_axial_vs_rotational_displacement}
	\end{figure}

The pressure gradient is of course readily available from the numerical simulation; the purpose of the model is to determine the source of this gradient.  According to equation \ref{eqn:p_transient}, the pressure downstream of a moving shock,  $p_2$, is a function of $M\text{n}^{\scales}_1$ and $p_1$. While $p_1$ is fixed by the upstream  flow conditions, $M\text{n}^{\scales}_1$ can be highly affected by the displacement of the shock wave due to the shock propagation velocity S. Figure~\ref{fig:illustration_axial_vs_rotational_displacement}  illustrates the impact of the shock displacement on S, which is given as a TOS output parameter. While an orthogonal shock displacement results in an uniform S distribution (figure~\ref{fig:illustration_axial_vs_rotational_displacement}(a)), a rotational shock displacement leads into a gradient in S along the shock wave (figure~\ref{fig:illustration_axial_vs_rotational_displacement}(a)). Here we seek to test whether the rotation of the shock is sufficient to explain the strength of the gradient observed in the data.

	\subsubsection*{Evolution of the pressure distribution downstream of the reflected shock}
	
\begin{figure}
		\centering
		\includegraphics[width=1\textwidth]{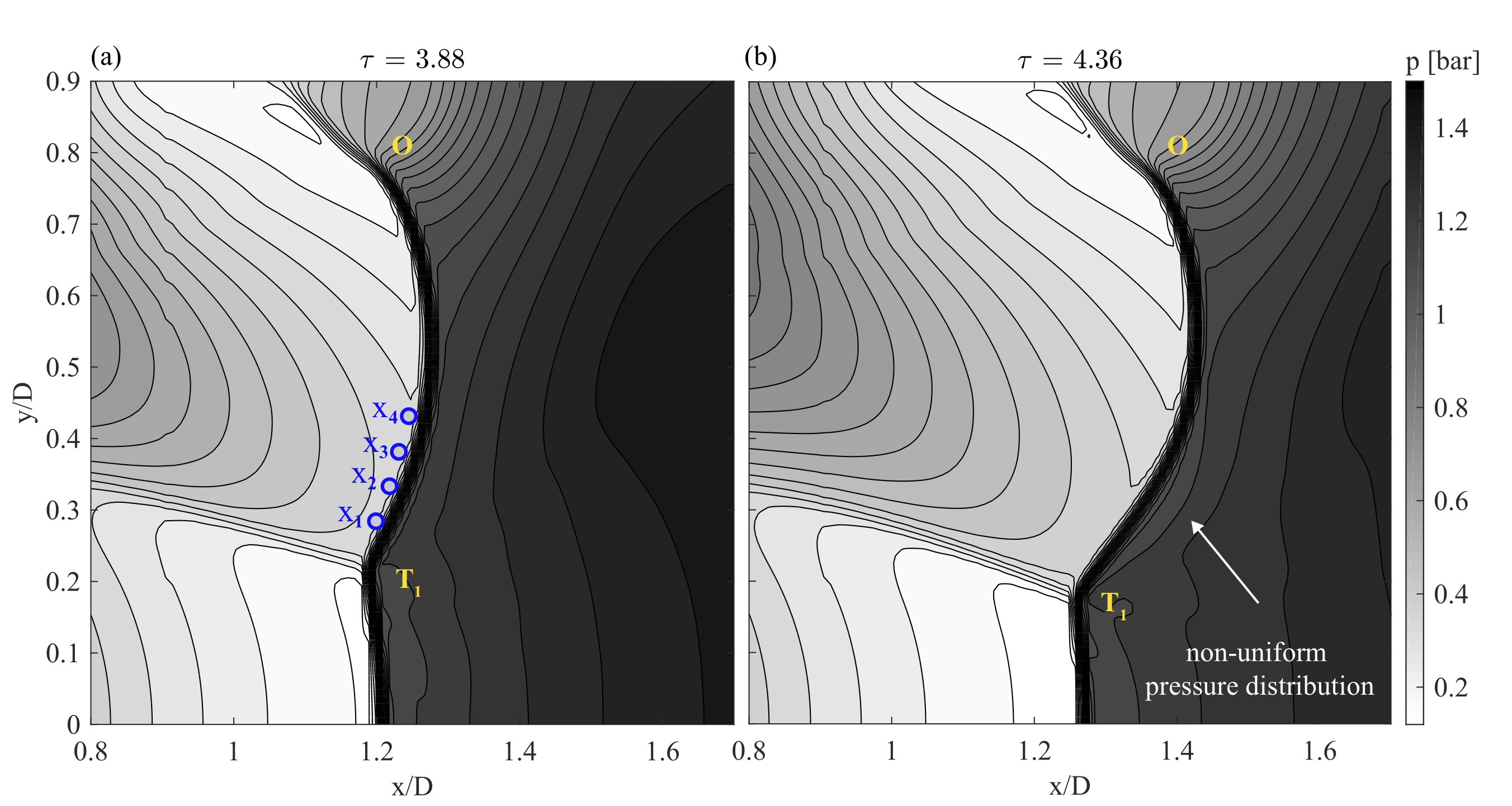}
		\caption{(a)Pressure distribution at \tFuenfUndZwanzig as derived from the numerical simulations at constant inflow conditions with \mbox{Ms = 1.71}. $\text{T}_1$ and O represents the first triple point and the tail of the VRES.. The TOS model is applied for $x_1$ to $x_4$. The result of the TOS analysis is given in table \ref{tab:p_dist_2}. (b) Numerical pressure distribution at \tSiebenUndZwanzigY, showing a pressure gradient on the downstream face of the reflected shock.}
		\label{fig:p_dist_2}
	\end{figure}

The pressure distribution downstream of the reflected shock is uniform up to \tZweiUndZwanzig (figure~\ref{fig:k20_p_an_uniform1}(b)), since the motion of the reflected shock at this time point is primarily translation rather than rotation; this translational motion is indicated by the constant S distribution in table~\ref{tab:p1_k20_p_an_uniform}. However, as the vortex ring and its embedded shock translate further downstream (t > \tZwanzigY), and pass the Mach disk (figure \ref{fig:p_dist_2}(a)), the inner part of the reflected shock  begins to tilt. The TOS model is applied to 4 discrete points upstream of the tilted part of the reflected shock, as shown in figure~\ref{fig:p_dist_2}(a). The results of the TOS analysis for \mbox{$x_1$ to $x_4$} are shown in table~\ref{tab:p_dist_2}. The model suggests a decreasing downstream pressure $p_2$ from \mbox{$x_1$ $\rightarrow$ $x_4$}.   Figure~\ref{fig:p_dist_2}(b) exhibits the pressure field obtained from the numerical simulation a short time later, at \tSiebenUndZwanzigY.  In accordance with the results of the TOS analysis, a non-uniform pressure distribution can be observed downstream of the reflected shock. 

% k20 - c 1 keynote
	\begin{table}
		\begin{small}
			\caption{TOS results for \tFuenfUndZwanzigY, shown in figure~\ref{fig:p_dist_2}(a). \label{tab:p_dist_2}}
			%        \begin{ruledtabular}
			\begin{tabular}{lcccccccccc}
				&   \multicolumn{3}{c}{input}   &  &                \multicolumn{4}{c}{output}                 &  &  \\ \cline{2-4}\cline{6-11}
						& $M_1$ 	& $p_1[bar]$			 & $\alpha$	 			&  & 		$M\text{n}^{\scales}_1$  			&			 $M^{\scales}_1$ &		 $\beta^{\scales} $ 	&	 $S[m/s]$ &					   $\frac{p_2}{p_1}$		  &  $p_2 [bar]$  \\ \midrule
				{$\boldsymbol x_1$} & 2.33  &    0.37  		  &   -1.3   		&  &       1.81               		&      1.97           		&   66.6     	    &    123   		 &      				3.65  		 & \textbf{1.35} \\
				{$\boldsymbol x_2$}  & 2.36  &    0.36    		&   0.2   		 &   &       1.77               	  &      1.97         		 &  64.3   			 &    134 		  &      				3.51 		   & \textbf{1.27} \\
				{$\boldsymbol x_3$} & 2.40  &    0.35    		&   2.4    		&   &       1.73               		 &      1.97           		&  61      		 &    149  		  &     				3.33 		   & \textbf{1.16} \\
				{$\boldsymbol x_4$}  & 2.44  &    0.33   		 &   4.2    	&   &       1.69              		  &      1.99           	&  58.3    		 &    161  		  &  					3.18  		  & \textbf{1.06} \\
		\end{tabular}
	\end{small}
\end{table}
	
	The aforementioned results indicate that the pressure gradient observed in the simulation data can be caused purely by the rotation of the oblique shock. The next step is to determine a more exact mechanism. Therefore, we consider next the spatial distribution of the flow between the jet core and the jet shear layer ($x_1$ $\rightarrow$ $x_4$ in figure \ref{fig:p_dist_2}(a)). According to table~\ref{tab:p_dist_2} there is a declining pressure ratio $\frac{p_2}{p_1}$ from \mbox{$x_1$ $\rightarrow$ $x_4$}, as $\frac{p_2}{p_1}(x_1)>\frac{p_2}{p_1}(x_2)>\frac{p_2}{p_1}(x_3)>\frac{p_2}{p_1}(x_4)$, which results in a pressure gradient downstream of the shock wave. According to equation \ref{eqn:p_transient}, the pressure ratio $\frac{p_2}{p_1}$ is a function of $M^{\scales}_1$  and $\sin \beta^{\scales}$. The term $\sin \beta^{\scales} $ can be linearized  to $ \beta^{\scales}$ under the small-angle approximation. Therefore, in the context of the model, the negative pressure ratio from \mbox{$x_1$ $\rightarrow$ $x_4$} can be ascribed to  either decreasing $M^{\scales}_1$, decreasing $\beta^{\scales}$, or both. While $M^{\scales}_1$ remains almost uniform from \mbox{$x_1$ $\rightarrow$ $x_4$}, a significant decrease  for $\beta^{\scales}$ is evident, as shown in table~\ref{tab:p_dist_2}.  Consequently, if the mechanism leading to the distribution of $M^{\scales}_1$  and $\beta^{\scales}$ is known, the formation of the non-uniform pressure region downstream of the reflected shock can likewise be determined.
	
	To elucidate the mechanism responsible for the distribution of $M^{\scales}_1$  and $\beta^{\scales}$, we consider the displacement of the reflected shock; the corresponding shock propagation velocity $S$ is given in table~\ref{tab:p_dist_2}. The  significant increase of the shock velocity from \mbox{$x_1$ $\rightarrow$ $x_4$}  indicates a strong tilting motion of the shock wave. The increase in the shock velocity $S$ leads inherently to a decrease in the relative Mach number in shock reference $M^{\scales}_1$, as $M^{\scales}_1 = M_1 - \frac{S}{a(x)}$. This correlation can also be recognized visually from figure~\ref{fig:prep_foot_sketch}, which illustrates a tilting shock wave. As shown in table~\ref{tab:p_dist_2}, the approaching Mach number $M_1$ increases from \mbox{$x_1$ $\rightarrow$ $x_4$},  which has the opposite effect on $M^{\scales}_1$, as can be seen from the equation above. However, the uniform distribution of $M^{\scales}_1$  over \mbox{$x_1$ $\rightarrow$ $x_4$} indicates, that the increase of $S$  compensates for the increase of $M_1$.  Consequently, the uniform distribution of $M^{\scales}_1$ is caused by the pronounced increase of $S$, i.e., due to the strong tilting of the reflected shock.  Moreover, the tilting motion of the reflected shock also affects the shock angle $\beta$. The rotation of the shock results  in an inherent reduction of the shock angle ($\beta^{\scales} < \beta$) as can be seen in figure \ref{fig:prep_foot_sketch}. As shown in table~\ref{tab:p_dist_2}, $\beta^{\scales}$  decreases from \mbox{$x_1$ $\rightarrow$ $x_4$}. Consequently, the tilting shock wave results in a negative pressure gradient by reducing $M^{\scales}_1$ and $\beta^{\scales}$ from \mbox{$x_1$ $\rightarrow$ $x_4$}. Additionally, a small decrease in upstream pressure $p_1$ and an increase in $\alpha$ supports the formation of the pressure gradient by simply decreasing $\beta$ and therefore $\beta^{\scales}$ (figure \ref{fig:prep_foot_sketch}). Hence, these results suggest that the tilting motion of the reflected shock and the alteration of the flow angle upstream of this shock are the primary mechanisms responsible  for the reduction of $M^{\scales}_1$ and $\beta^{\scales}$, and thereby for the formation of the pressure gradient downstream of the reflected shock.

	At later times, the vortex ring propagates further downstream (see figures~\ref{fig:p_dist_3} and \ref{fig:p_dist_4}). Thus the disparity in the angle dictated by the first triple point and that required by the VRES increases. Hence, a kink forms gradually within the shock wave (figures~\ref{fig:p_dist_3} and \ref{fig:p_dist_4}), separating the reflected shock and the VRES.  A dotted line in figures~\ref{fig:p_dist_3} and \ref{fig:p_dist_4}, originating from K separates two regions A and B, downstream of the reflected shock and the VRES respectively. The TOS model is applied to three points for each region (figure \ref{fig:p_dist_3} and \ref{fig:p_dist_4}). The corresponding results are presented in table~\ref{tab:p_dist_3} and \ref{tab:p_dist_4}. We first evaluate the results for region A, before proceeding further with region B.  The comparison of the CFD pressure distribution at \tFuenfUndZwanzig to \tFuenfUndDreissigY, shown in figures \ref{fig:p_dist_2} - \ref{fig:p_dist_4}, exhibits an increases of the gradient in \NPG region with time. In conformity with the CFD pressure distribution, the TOS results predict an increase of the pressure gradient with time downstream of the reflected shock; the pressure gradient $p_2(x_1)/p_2(x_4)$ increases by approximately 8 \% between \tFuenfUndZwanzig and \tAchtUndZwanzig, and by 86\% from \tAchtUndZwanzig to \tdreiunddreissig (tables~\ref{tab:p_dist_2} to \ref{tab:p_dist_4}). A comparison of the shock velocity S indicates that the tilting motion of the reflected shock becomes significantly stronger as $ S(x_4)/S(x_1)$ increases with time. The quantity $  S(x_4)/S(x_1) $ increases by approximately 26 \% from \tFuenfUndZwanzig to \tAchtUndZwanzig and  1800\% from \tAchtUndZwanzig to \tDreissig (tables~\ref{tab:p_dist_2} to \ref{tab:p_dist_4}). For the reasons indicated above, the strong tilting results in a stronger compression, as \mbox{$ \Pi_{\tDreissigX} > \Pi_{\tAchtUndZwanzigX} > \Pi_{\tFuenfUndZwanzigX}$}, where $\Pi = \frac{p_2}{p_1}(x_1)/\frac{p_2}{p_1}(x_4)$. Hence, the gradient of the \NPG region increases with time. The tilting of the shock is not the only mechanism by which the \NPG changes as a function of time: there is also a small increase in the pressure gradient upstream of the reflected shock $p_1(x_1)/p_1(x_4)$ of 11, 17 and 30\%  from \mbox{$x_1$ $\rightarrow$ $x_4$} between \, \tAchtUndZwanzig and \tDreissig. Similarly, there is an increase in the flow angle $\alpha$ of 5.5$^\circ$, 8.3$^\circ$ and 16.6$^\circ$, respectively.  Nevertheless, the contribution of these mechanisms is relatively small compared to the gradients induced by the motion of the shock; the tilting of the reflected shock is the main reason for a pronounced pressure gradient downstream of the reflected shock wave. 

	\begin{table}
		\begin{small}
			\caption{TOS results for \tAchtUndZwanzigY, shown in figure~\ref{fig:p_dist_3}(a). \label{tab:p_dist_3}}
			%        \begin{ruledtabular}
			\begin{tabular}{lcccccccccc}
				&   \multicolumn{3}{c}{input}   &  &                \multicolumn{4}{c}{output}                &                   &               \\ \cline{2-4}\cline{6-11}
						& $M_1$ & $p_1[bar]$	 & $\alpha$ &  	&			 $M\text{n}^{\scales}_1$ & $M^{\scales}_1$	 &		 $\beta^{\scales} $ & $S[m/s]$ & $\frac{p_2}{p_1}$ &  $p_2 [bar]$  \\ \midrule
				{$\boldsymbol x_1$} & 2.23  &    0.42    &   -7.9   &  &       1.74          &      2.04       &   58.5   &   70    &      3.35       & \textbf{1.40} \\
				{$\boldsymbol x_2$} & 2.29  &    0.41    &   -4.3   &  &       1.68           &      2.07       &   54.1   &   81    &       3.13        & \textbf{1.27} \\
				{$\boldsymbol x_3$} & 2.37  &    0.37    &   -1.0   &  &       1.64         &      2.14       &   49.9   &   91    &      2.95       & \textbf{1.09} \\
				{$\boldsymbol x_4$} &  2.4  &    0.35    &   0.4    &  &       1.61            &      2.17       &  48.1          &   95   &      2.87     & \textbf{1.02} \\ 
				\hline
				{$\boldsymbol x_5$} & 2.57  &    0.29    &   5.1    &  &       1.96          &      2.00       &   78.1   &   169    &       4.31        & \textbf{1.24} \\
				{$\boldsymbol x_6$} & 2.63  &    0.27    &   7.1    &  &       1.97           &      2.04       &   75.3   &   177   &       4.36        & \textbf{1.16} \\
				{$\boldsymbol x_7$} & 2.73  &    0.23    &   10.2   &  &       2.06            &      2.07       &   83.5   &   185    &      4.79      & \textbf{1.12}
			\end{tabular}
		\end{small}
	\end{table}

	As seen by the TOS model and the numerical simulations both the size and strength of the \NPG region grow with time. The flow evolution from \tAchtUndZwanzig to \tDreissig (figure~\ref{fig:p_dist_3}) shows that the reflected shock ($\text{T}_1$-K) elongates with time. Consequently, the \NPG region covers a wider area downstream of the reflected shock, as shown in figure \ref{fig:p_dist_3}(b). However, the \NPG region occurs only downstream of the reflected shock, from $\text{T}_1$ to K, and ends at the  A-B-interface. These observations based on the CFD results agree again with the results from the TOS analysis, shown in table~\ref{tab:p_dist_3};  the \NPG region elongates from $x_1$ to $x_4$  (figure~\ref{fig:p_dist_3}(b)), as the pressure $p_2$ decreases from \mbox{$x_1$ $\rightarrow$ $x_4$},  but there is a positive pressure gradient from A to B at their interface, as $p_2(x_5) > p_2(x_4)$ shown in table~\ref{tab:p_dist_3}. This positive pressure gradient is the origin of a new shock wave, the shocklet, as will be discussed in the following. 

\subsubsection*{Evolution of the pressure distribution downstream of the VRES}

\begin{figure}
		\centering
		\includegraphics[width=1\textwidth]{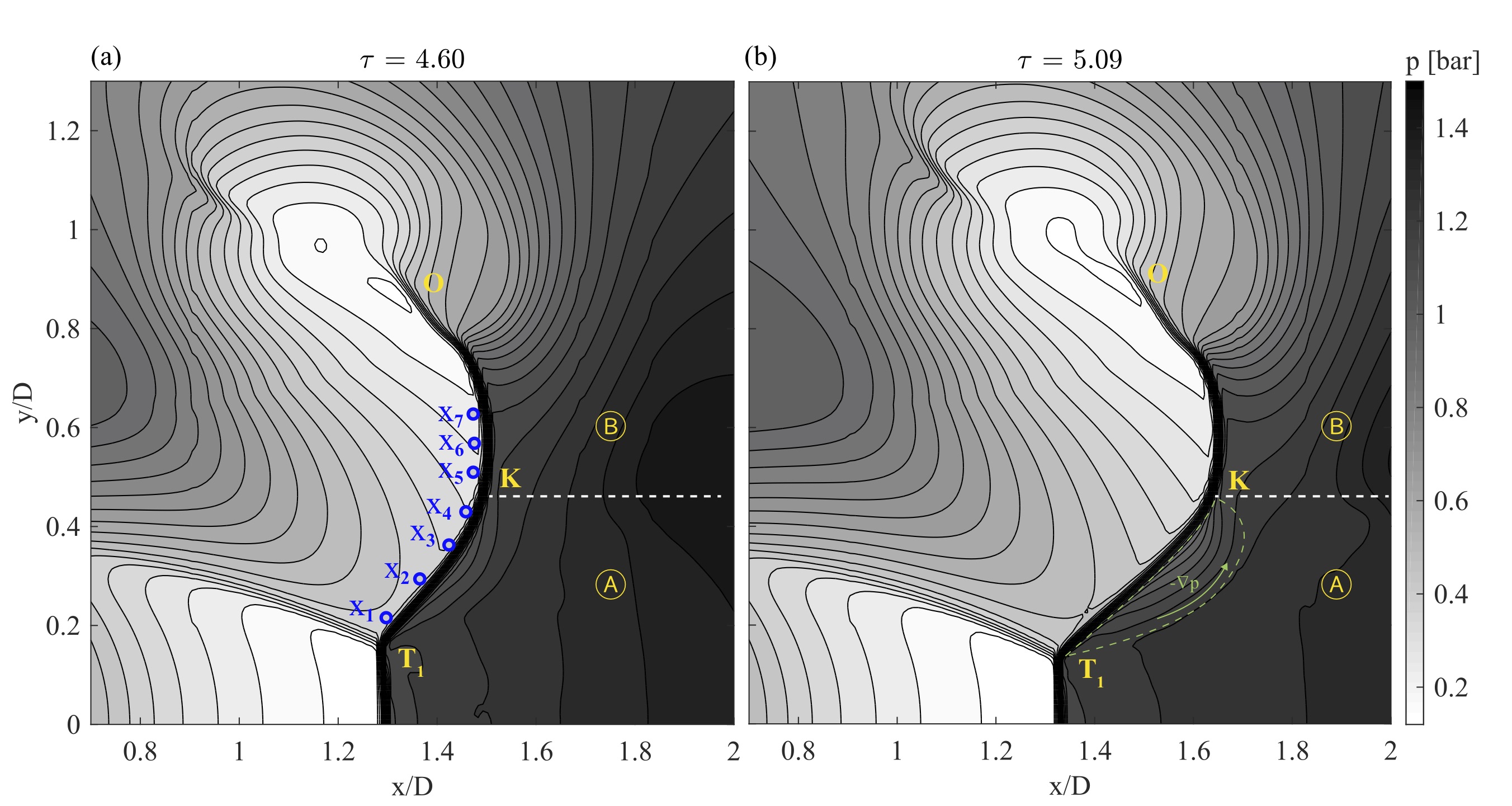}
		\caption{(a)pressure distribution at \tAchtUndZwanzig as derived from the numerical simulations at constant inflow conditions with \mbox{Ms = 1.71}. $\text{T}_1$ and O represents the first triple point and the tail of the VRES. The  TOS model is applied for $x_1$ to $x_7$. The result of the TOS analysis is given in table \ref{tab:p_dist_3}. (b) pressure distribution at \tDreissig.}
		\label{fig:p_dist_3}
	\end{figure}

	The formation of a positive pressure gradient from A to B can be further examined by considering the TOS results for the region B given in table~\ref{tab:p_dist_3}. As shown in figure~\ref{fig:p_dist_3} and also  indicated by the shock velocity $S$ for $x_5$ to $x_7$ in table~\ref{tab:p_dist_3}, the VRES tilts barely, but translates predominantly in the axial direction. This is also the case for the approaching flow in region B, indicated by small $\alpha$  for $x_5$ to $x_7$ in table~\ref{tab:p_dist_3}. The combination of the vertical shock, moving in the axial direction and small $\alpha$ results in significantly large shock angles $\beta$, leading to high-pressure ratios \mbox{$\frac{p_2}{p_1}$}  in region B ($x_5$ to $x_7$ in table~\ref{tab:p_dist_3}). Hence, the pressure in region B is higher than A in the vicinity of their interfaces; there is a positive pressure gradient from A to B.

	\begin{table}
		\begin{small}
			\caption{TOS results for \tdreiunddreissigY, shown in figure~\ref{fig:p_dist_4}(a) . \label{tab:p_dist_4}}
			%        \begin{ruledtabular}
			\begin{tabular}{lcccccccccc}
				&   \multicolumn{3}{c}{input}   &  &                \multicolumn{4}{c}{output}                 &  &  \\ \cline{2-4}\cline{6-11}
						& $M_1$ 		& $p_1[bar]$ 		&	 $\alpha$ 		&  &		 $M\text{n}^{\scales}_1$ & 		$M^{\scales}_1$ 		&	 $\beta^{\scales} $ & $S[m/s]$ &   $\frac{p_2}{p_1}$  				&  $p_2 [bar]$  \\ \midrule
				{$\boldsymbol x_1$} & 2.09   &   0.46          &   -14.3   &    &       1.61               &      2.09             &   50.6       &   2     &             2.87             & \textbf{1.33} \\
				{$\boldsymbol x_2$} & 2.05   &   0.59          &   -9.2    &    &       1.40                 &      2.01             &   44.3       &   20     &             2.13             & \textbf{1.26} \\
				{$\boldsymbol x_3$}  & 2.31   &   0.42          &   -2.1    &    &         1.31              &      2.24           &   35.9		   &   39     &           1.84             & \textbf{0.78} \\
				{$\boldsymbol x_4$}  & 2.52   &  0.32          &   2.3    &    &         1.24                &      2.43           &     30.6      &   51    &             1.62             & \textbf{0.52} \\
				\hline
				{$\boldsymbol x_5$}  & 2.65   &  0.27          &   4.7    &    &        2.06               &      2.08          &     82.6     &   168   &           4.81            & \textbf{1.28} \\
				{$\boldsymbol x_6$}  & 2.73   &  0.24          &   6.6    &    &        2.09              &     2.12           &     80.0     &   178   &         4.91            & \textbf{1.18} \\
				{$\boldsymbol x_7$}  & 2.79   &  0.22          &   8.3    &    &         2.10              &      2.15           &     77.7     &   186  &          4.98           & \textbf{1.09} \\
				
			\end{tabular}
		\end{small}
	\end{table}
            
\begin{figure}
		\centering
		\includegraphics[width=1\textwidth]{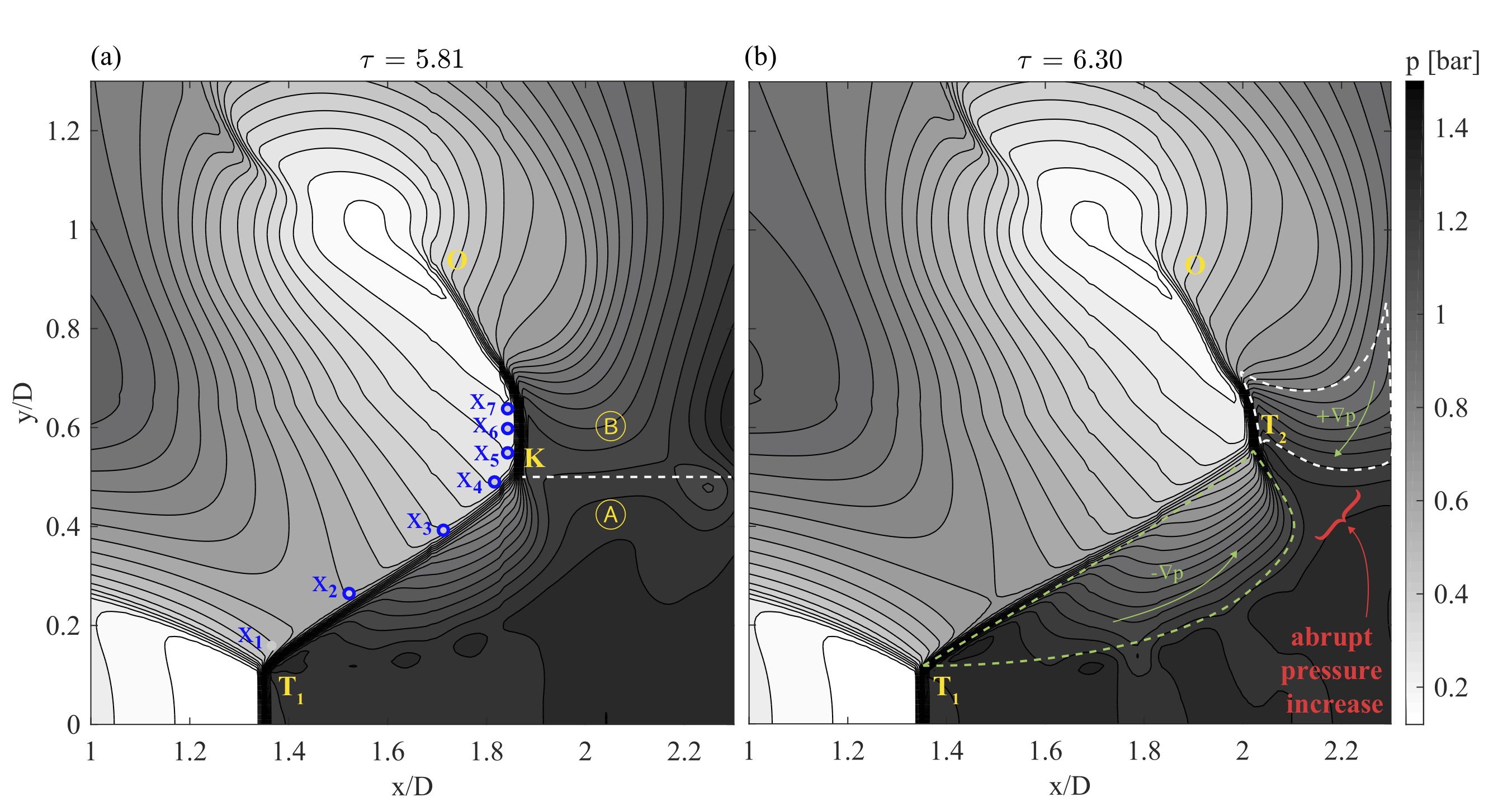}
		\caption{(a)pressure distribution at \tdreiunddreissig as derived from the numerical simulations at constant inflow conditions with \mbox{Ms = 1.71}. $\text{T}_1$ and O represents the first triple point and the tail of the VRES. The TOS model is applied for $x_1$ to $x_7$. The result of the TOS analysis is given in table \ref{tab:p_dist_4}. (b) pressure distribution at \tFuenfUndDreissig.}
		\label{fig:p_dist_4}
	\end{figure}

	Figure~\ref{fig:p_dist_4}(a) shows the flow evolution at a later stage in time for \tdreiunddreissig and \tFuenfUndDreissigY. The corresponding TOS results for $x_1$ to $x_7$ at \tdreiunddreissig are given in table~\ref{tab:p_dist_4}. The evolution of the \NPG region can be evaluated for an extended period of time based on the pressure distribution \tAchtUndZwanzig to \tFuenfUndDreissig shown in figures~\ref{fig:p_dist_3} and \ref{fig:p_dist_4}. It is evident that the \NPG  region enlarges further and its pressure gradient increases with time.  Similar to the \NPG, also the pressure gradient in region B becomes more distinctive with time. The TOS results confirm again the CFD results, showing an increase in pressure gradient with time in both regions A and B, as \mbox{$\big[p_2(x_1)/p_2(x_4)\big]_{\tdreiunddreissigX} > \big[p_2(x_1)/p_2(x_4)\big]_{\tAchtUndZwanzigX} > \big[p_2(x_1)/p_2(x_4)\big]_{\tFuenfUndZwanzigX}$} and  \mbox{$\big[p_2(x_5)/p_2(x_7)\big]_{\tdreiunddreissigX} > \big[p_2(x_5)/p_2(x_7)\big]_{\tAchtUndZwanzigX}$}. Both, the CFD and the TOS results show the pressure gradients in both regions A and B increase with time. The pressure gradient downstream of the reflected shock in region A, from $\text{T}_1$ to K, is negative.  In contrast,  there is a positive pressure gradient in region B, from O to K, as marked in figure~\ref{fig:p_dist_4}(b). Hence, the increase in the pressure gradient in A and B results in higher pressure ratio along the  A-B-interface. Based on the TOS results, shown in (table~\ref{tab:p_dist_3} and \ref{tab:p_dist_4}), the pressure ratio $p_2(x_5)/p_2(x_4)$     increases from \tAchtUndZwanzig to \tdreiunddreissig by \mbox{102 \%}. The result of this evolution can be observed in figure \ref{fig:p_dist_4}(b). The negative pressure gradient from T to K and the positive pressure gradient from O to K lead to an increase of the pressure within a very small region, between the $- \nabla p$ and $+ \nabla p$ regions (figure~\ref{fig:p_dist_4}(b)). As both pressure gradients intensify with time, an abrupt pressure change occurs at the intersection of these regions. Consequently, the abrupt pressure rise leads to the formation of a new shock wave.
	
	Now, the formation of the second triple point and the shocklet  can be summarized. The abrupt pressure rise, which necessitates the formation of the shocklet, is induced by the evolution of the pressure distribution downstream of the reflected shock.  The pressure downstream of the reflected shock is highly affected by the displacement of this shock wave over time. This is due to two main facts; firstly, the flow velocity upstream of the reflected shock is higher than the propagation velocity of the shock wave. Consequently, the flow downstream of the shock wave is driven by the  prior motion of this shock wave. Secondly, the displacement of the reflected shock is non-uniform along the shock wave with the reflected shock tilting toward the jet center line,  driven by the convection of the vortex ring. As the vortex ring moves further away from the first triple point, the angle dictated by the first triple point and the one required from the part of the shock wave, which is embedded in the vortex ring (VRES), differ. Hence, a kink appears within the shock wave, which separates the reflected shock from the VRES. Due to the rotational motion of the reflected shock, a negative pressure gradient arises in the radial direction, from the jet core to the jet boundary. This pressure gradient increases with time, as the reflected shock extends and rotates further. Hence, the pressure becomes relatively low downstream of the reflected shock. Its minimum value occurs right below the kink. In contrast, the pressure downstream of the VRES is relatively high. This is mainly due to the nearly axial displacement of the shock wave, leading into large shock angles along the  wave. Consequently, the pressure above the kink becomes much higher than below the kink, resulting in an abrupt pressure rise. As the abrupt pressure rise leads into the formation of a new shock wave (shocklet), the kink becomes a triple point. Finally, the shocklet, the reflected shock and the VRES forms the second triple point configuration of the transient supersonic starting jet.

	\section{conclusion}
	The dynamic evolution of a starting transient supersonic flow has been studied by utilizing numerical simulations and high-resolution high-speed schlieren measurements. It has been shown that for a sufficiently strong leading shock, the interaction of the secondary shock system with the VRES will result in the formation a second triple point. Experimental evidence is provided for the presence of a second triple shock configuration along with a shocklet between the reflected shock and the slipstream, which results in the formation of further KH vortices. 
	
	A simple model was developed based on 1D shock relations, in an attempt to determine the source of pressure distributions in the flow which could give rise to the shocklet. A comparison of the output of this model to the results of the numerical simulations suggested that  the shocklet forms due to a different mechanism than the classical Mach reflection responsible for the first triple point. 
	
	The formation of the second triple point is initiated by the transient motion of the reflected shock, which is induced by the convection of the vortex ring. As the vortex ring overtakes the Mach disk, the part of the reflected shock next to the core begins to tilt, while the outer part of the shock propagates almost uniformly further downstream. Consequently, a kink appears in the reflected shock, separating the reflected shock from the vortex ring embedded shock. Downstream of the reflected shock a negative pressure gradient in radial direction occurs, which is caused by the rotational motion of the reflected shock wave. This pressure gradient region grows in size and strength, as the reflected shock elongates and rotates further. Hence, the pressure just below the kink decreases with time. In contrast, the pressure downstream of the vortex ring embedded shock, particularly in the vicinity of the kink, is relatively high.  Therefore, an abrupt pressure rise along the kink takes place. The kink becomes a triple point, while the abrupt pressure rise results in the formation of a new shock wave. 
	
	\FloatBarrier%-------------------------------------------------------------------
	
	\begin{acknowledgments}
		The authors gratefully acknowledge support by the Deutsche Forschungsgemeinschaft (DFG) as part of the Collaborative Research Center SFB 1029 "Substantial efficiency increase in gas turbines through direct use of coupled unsteady combustion and flow dynamics". The authors also thank Dr.\ N.\ Nikiforakis' team at the Laboratory for Scientific Computing, Cambridge University, for the generous access to LSC\_AMR, their cartesian grid--cut cell--compressible flow solver. The authors also gratefully acknowledge support by the ARC DP190102220. 
	\end{acknowledgments}
	
	\input{0_shocklet_PRF_arxive.bbl}
%	\bibliography{0_shocklet_PRF}
	
\end{document}